\documentclass[%
 reprint,
 superscriptaddress,
 amsmath,%amssymb,
 aps,
 prb
]{revtex4-2}

\usepackage[pdftex]{graphicx, color}% Include figure files
\usepackage[pdftex,
            colorlinks=true,
            citecolor=blue,
            linkcolor=blue]{hyperref}
\usepackage{newtxtext, newtxmath}
\usepackage{bm}% bold math
\usepackage{braket, mathtools}
\usepackage{bbm}
\usepackage{ulem}
\usepackage{tabularx}
\graphicspath{{./images/}}

\begin{document}

\title{Nonlinear optical responses in noncentrosymmetric superconductors}

\author{Hiroto Tanaka}
\email[]{tanaka.hiroto.54z@st.kyoto-u.ac.jp}
\affiliation{%
Department of Physics, Graduate School of Science, Kyoto University, Kyoto 606-8502, Japan
}%

\author{Hikaru Watanabe}
\affiliation{%
%  RIKEN Center for Emergent Matter Science (CEMS), Wako 351-0198, Japan
Research Center for Advanced Science and Technology, University of Tokyo, Komaba Meguro-ku, Tokyo 153-8904, Japan
}%

\author{Youichi Yanase}
\affiliation{%
Department of Physics, Graduate School of Science, Kyoto University, Kyoto 606-8502, Japan
}%

\date{\today}

\begin{abstract}
The unique nonreciprocal responses of superconductors, which stem from the Cooper pairs’ quantum condensation, have been attracting attention. Recently, theories of the second-order nonlinear response in noncentrosymmetric superconductors were formulated based on the Bogoliubov-de Gennes theory. In this paper, we study the mechanism and condition for second-order optical responses of time-reversal symmetric superconductors. The numerical results show the characteristic photocurrent and second harmonic generation in the superconducting state. However, the superconductivity-induced nonlinear optical responses disappear under some conditions on pair potential. We show that the coexistence of intraband and interband pairing is necessary for the second-order superconducting optical responses. In addition, the superconducting Berry curvature factor, which is related to a component of Berry curvature in the superconducting state, is essential for the nonlinear responses. Thus, we derived the microscopic conditions where the superconducting nonlinear response appears.
\end{abstract}

\maketitle

\section{Introduction}
Various optical measurements have been used to reveal the complex ordered states of quantum materials. Especially, nonlinear responses yield rich information such as symmetry and geometric properties of quantum phases~\cite{Orenstein2021}. For example, the second-order optical response is a useful tool for probing the microscopic parity breaking in complex ordered states because the second-order response requires broken inversion symmetry due to the symmetry constraint~\cite{Zhao2017,Torre2021}. Furthermore, in the application field, the nonlinear response is expected to be a new basis of the photo-electron converter. Therefore, nonlinear optics in quantum materials have been attracting attention from fundamental science to technology due to their various applications.

The leading nonlinear response is represented by the second-order optical response. The formula of second-order nonlinear response is generally given by
\begin{equation}
\label{eq:2nd_nonlinear_res}
\left<\mathcal{J}^{\alpha}(\omega)\right>_{(2)}=\int\frac{d\Omega}{2\pi}\sigma^{\alpha;\beta\gamma}(\omega;\Omega,\omega-\Omega)E^{\beta}(\Omega)E^{\lambda}(\omega-\Omega).
\end{equation}
In particular, second-order optical responses for a monochromatic light are comprised of the second harmonic generation and the photogalvanic effect, which are denoted by $\sigma(2\omega;\omega,\omega)$ and $\sigma(0;\omega,-\omega)$, respectively.

The second harmonic generation, which is a frequency doubling of the light through interaction with media, is particularly sensitive to the microscopic space inversion parity ($\mathcal{P}$) breaking in materials. In the multiferroic materials science, indeed, the second harmonic generation has been introduced as a probe for magnetic structures~\cite{Fiebig2005}. Moreover, the second harmonic generation experiments have revealed exotic states breaking the symmetry in correlated electron systems such as cuprate high-temperature superconductors~\cite{Zhao2017,Torre2021}.

The photogalvanic effect, which is a photo-induced direct current (photocurrent), is also influenced by various properties of the system such as spatially-inhomogeneous and asymmetric structures. Interestingly, the intrinsic photogalvanic effect is related to the geometric property of the Bloch wave function. Therefore, topological materials are potential candidate systems for a new dc photo-electric converter~\cite{Liu2020,Orenstein2021}. In addition, it has been shown that time-reversal ($\mathcal{T}$) symmetry breaking may enhance the photogalvanic effect~\cite{Watanabe2021,Ahn2020}. Indeed, parity-breaking magnets are expected to be tunable photo-electric converters due to the switchable magnetic order~\cite{Ogawa2016,Burger2020}.
% A recent study proposed that the photocurrent can be produced by the light without any optical absorption in metals breaking both $\mathcal{P}$ and $\mathcal{T}$ symmetries~\cite{Onishi2022}.

It has been widely known that superconductors show remarkable electromagnetic properties such as zero resistivity and the Meissner effect.
Superconductors also host unique nonlinear and nonreciprocal responses stemming from the Cooper pairs’ quantum condensation. For example, the Higgs mode, the amplitude mode of the order parameter, %is the important property of the superconductor. It is reported that the frequency dependence of 
gives rise to the third-order optical response %can probe the Higgs mode
\cite{Matsunaga2013,Cea2016}. In addition, the superconducting fluctuation~\cite{Wakatsuki2017,Wakatsuki2018} and the dynamics of vortices~\cite{Hoshino2018,Ideue2020} enhance the nonreciprocal electric transport, such as the magneto-chiral anisotropy. Recently, a superconducting diode effect, which means zero resistivity in a current direction while a finite resistivity in the opposite direction, 
has been observed in a noncentrosymmetric superconductor~\cite{Ando2020} and intensively studied theoretically~\cite{Daido2022,Yuan2022,He2022}.

From the above perspectives, in this paper, we focus on the second-order nonlinear optical response of noncentrosymmetric superconductors. The noncentrosymmetric superconductors have been attracting attention as a platform for exotic quantum phenomena such as the mixed singlet and triplet pairing~\cite{Gor'kov2001,Frigeri2004,Bauer2012,Smidman2017} and topological superconductivity~\cite{Alicea2012,Sato2016}. %and the anomalous paramagnetic effect. 
For example, the characteristic spin-momentum-locked electronic structure may boost the upper critical field beyond the Pauli-Clogston-Chandrasekhar limit~\cite{Bauer2012,Kimura2007,Settai2008}. In particular, Ising superconductivity has been reported in transition-metal dichalcogenides, such as gated $\mathrm{MoS_2}$~\cite{Saito2016,Lu2015}, $\mathrm{NbSe_{2}}$~\cite{Xi2016}, and $\mathrm{TaS_{2}}$~\cite{delaBarrera2018}, in which the broken inversion symmetry leads to a large spin-orbit coupling and huge upper critical field.
%such as $\mathrm{NbSe_{2}}$, $\mathrm{TaS_{2}}$ \cite{Xi2016,delaBarrera2018}, $\mathrm{TaSe_2}$ \cite{Yokota2000}, and . 
Recently, the effect of parity breaking was observed in the nonlinear optics of an $s$-wave superconductor thin film, 
where the $\mathcal{P}$ symmetry is broken not by the crystalline structure but by the supercurrent~\cite{Nakamura2020}. Thus, the second-order nonlinear response in superconductors is developing, and theoretical studies are desirable. In recent theories, the second-order nonlinear response was formulated based on the Bogoliubov-de Gennes Hamiltonian~\cite{Xu2019,Watanabe2022}. However, it is still unclear how superconductivity influences nonlinear optical properties. In particular, the microscopic condition for superconducting nonlinear optical responses remains elusive. 
Thus, it is desirable to perform a comprehensive analysis of the superconducting nonlinear optical responses based on a canonical model.
Our goal is to elucidate the exotic properties of noncentrosymmetric superconductors by the nonlinear optics. 

\begin{figure*}[tbp]
 \includegraphics[width=\linewidth]{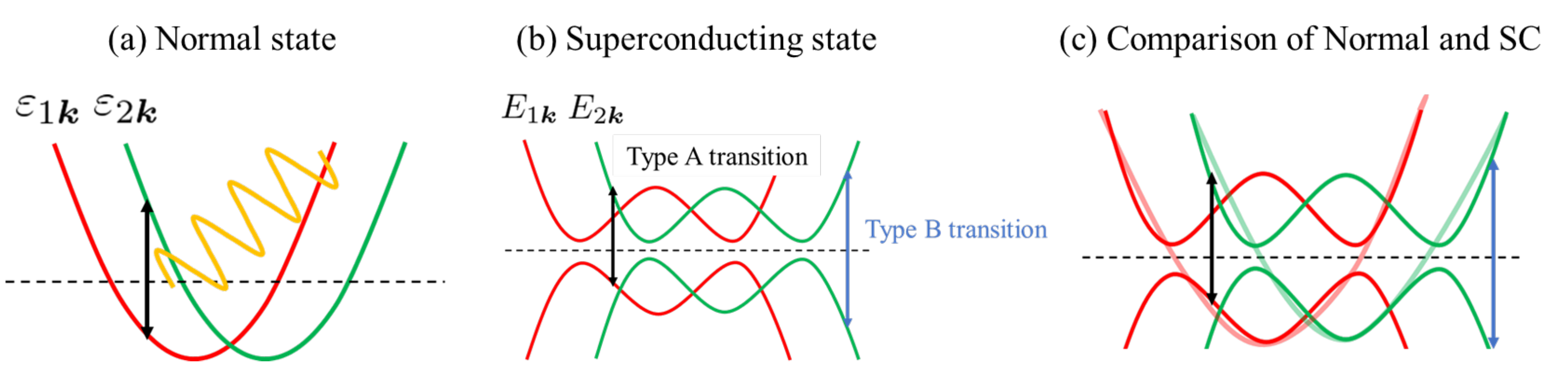}
 \caption{Schematic illustration of the transitions which contribute to the second-order nonlinear responses. $\varepsilon_{i}$ and $E_{i}$ are band energy in the normal state and positive band energy in the superconducting state, respectively. (a) In the normal state, $\varepsilon_{1} \leftrightarrow \varepsilon_{2}$ transition dominantly contributes to the second-order nonlinear responses. (b) In the superconducting state, the type A and type B transitions contribute to the second-order nonlinear responses. The type A is $E_{i}\leftrightarrow -E_{j} \quad (i\neq j)$ transition and the type B is $E_{i}\leftrightarrow -E_{i}$ transition. (c) Type A transition is similar to the $\varepsilon_{1} \leftrightarrow \varepsilon_{2}$ transition. Thus, the normal nonlinear response, which stems from the type A transition, appears even in the superconducting state. The type B transition contributes to the superconducting nonlinear response.}
 \label{fig:transition_image}
\end{figure*}

We investigate the second-order nonlinear responses of $\mathcal{T}$-symmetric superconductors with two-band model Hamiltonian. The photocurrent and second harmonic generation are demonstrated under various conditions on the pair potential. In this paper, we show that the coexistence of intraband and interband pairing is necessary for the second-order nonlinear responses unique to the superconducting state. The microscopic mechanism will be explained based on an illustration of Fig.~\ref{fig:transition_image}. In the superconducting state, the type A and type B transitions can contribute to the second-order nonlinear responses (Fig.~\ref{fig:transition_image}(b)). The normal nonlinear response mainly stems from the type A transition, which is similar to the transition in the normal state shown in Fig.~\ref{fig:transition_image}(a). On the other hand, the type B transition, which is unique to the superconducting state, contributes to the superconducting nonlinear response (Fig.~\ref{fig:transition_image}(c)). Our results show that the nonlinear response due to the type B transition is prohibited when the pair potential does not contain either the intraband or interband components. In addition, we introduce the superconducting Berry curvature factor that represents the relation between the $g$-vector of antisymmetric spin-orbit coupling and the $d$-vector of spin-triplet pairing. We demonstrate that the second-order nonlinear responses are suppressed when the superconducting Berry curvature factor vanishes. 

The outline of the paper is given below. In Sec.~\ref{sec:formulation}, we briefly explain the formulation of the nonlinear optical response in superconductors. We review the superconductivity-induced nonlinear response, which we call anomalous nonlinear response. In Sec.~\ref{sec:method}, the microscopic requirement in which the second-order nonlinear optical responses are suppressed is discussed based on the analytical results of the injection current. We show a conjecture that the coexistence of intraband and interband pairing and Berry curvature factors are essential for the second-order nonlinear optical responses. In Sec.~\ref{sec:result}, we verify the expectations and demonstrate the photocurrent and second harmonic generation characteristic of the superconducting state by numerical calculations. In Sec.~\ref{sec:discussion}, we summarize the results of Sec.~\ref{sec:result} and discuss tuning the photocurrent with varying the proportion of intraband and interband pairing. The contents are concluded, and future research is discussed in Sec.~\ref{sec:conclusion}.
Throughout this paper, we present formulas with the unit $\hbar=1$ (Dirac constant) and $q=1$ (electron charge).

\section{Formulation of nonlinear optical responses in superconductors}
\label{sec:formulation}
\subsection{General formula for nonlinear conductivity}
We consider a superconducting system described by the Bogoliubov-de Gennes (BdG) Hamiltonian~\cite{Watanabe2022}:
\begin{equation}
\mathcal{H}_{\mathrm{BdG}} = \frac{1}{2}\sum_{\bm{k}}\bm{\Psi_{k}}^{\dagger} H_{\mathrm{BdG}}({\bm{k}}) \bm{\Psi_{k}} + \mathrm{const},
\end{equation}
where Nambu spinor is $\bm{\Psi_{k}}^{\dagger} = (\bm{c}^{\dagger}, \bm{c}^{T})$ with the creation ($\bm{c}^{\dagger}$) and annihilation ($\bm{c}$) operators of electrons. The BdG Hamiltonian $H_{\mathrm{BdG}}$ consists of the normal-state Hamiltonian $H_{\mathrm{N}}(\bm{k})$ and the pair potential $\Delta_{\bm{k}}$,
\begin{equation}
H_{\mathrm{BdG}}(\bm{k}) = 
\begin{pmatrix}
H_{\rm N}(\bm{k}) & \Delta_{\bm{k}} \\
\Delta^{\dagger}_{\bm{k}} & -\left[H_{\rm N}(-\bm{k})\right]^{T} \\
\end{pmatrix}.
\end{equation}
With the velocity gauge $\bm{E}=-\partial_{t}\bm{A}(t)$, the couplings to external electric fields are introduced by the minimal coupling prescription $H_{\rm N}(\bm{k}) \rightarrow H_{\rm N}(\bm{k}-q\bm{A})$. Thus, the vector potential $\bm{A}$ dependence of the BdG Hamiltonian is given by 
\begin{equation}
H_{\mathrm{BdG}}(\bm{k},\bm{A}) = 
\begin{pmatrix}
H_{\rm N}(\bm{k}-q\bm{A}) & \Delta_{\bm{k}} \\
\Delta^{\dagger}_{\bm{k}} & -\left[H_{\rm N}(-\bm{k}-q\bm{A})\right]^{T} \\
\end{pmatrix}.
\end{equation}
Following the standard perturbative treatments, we evaluate the expectation value of the electric current density as 
\begin{equation}
\left<\mathcal{J}^{\alpha}(\omega)\right>=\sum_{n=1}\left<\mathcal{J}^{\alpha}(\omega)\right>_{(n)},
\end{equation}
where $\left<\mathcal{J}^{\alpha}(\omega)\right>_{(n)}$ is the electric current of the $n$-th order in $\bm{A}$.
The formula for the second-order nonlinear conductivity is given by
\begin{align}
&\sigma^{\alpha;\beta\gamma}(\omega;\omega_{1},\omega_{2}) \notag \\
&= \frac{1}{2(i\omega_{1} - \eta)(i\omega_{2} - \eta)}\sum_{\bm{k}}\Biggl[\sum_{a}\frac{1}{2}J^{\alpha\beta\gamma}_{aa}f_{a} \notag \\
&+ \sum_{a,b}\frac{1}{2}\left(\frac{J^{\alpha\beta}_{ab}J^{\gamma}_{ba}f_{ab}}{\omega_{2}+i\eta-E_{ba}} + \frac{J^{\alpha\gamma}_{ab}J^{\beta}_{ba}f_{ab}}{\omega_{1}+i\eta-E_{ba}}\right) \notag \\
&+ \sum_{a,b}\frac{1}{2}\frac{J^{\alpha}_{ab}J^{\beta\gamma}_{ba}f_{ab}}{\omega+2i\eta-E_{ba}} \notag \\
&+ \sum_{a,b,c} \frac{1}{2}\frac{J^{\alpha}_{ab}}{\omega+2i\eta-E_{ba}}\left(\frac{J^{\beta}_{bc}J^{\gamma}_{ca}f_{ac}}{\omega_2+i\eta-E_{ca}}-\frac{J^{\beta}_{ca}J^{\gamma}_{bc}f_{cb}}{\omega_2+i\eta-E_{bc}}\right) \notag \\
&+ \sum_{a,b,c} \frac{1}{2}\frac{J^{\alpha}_{ab}}{\omega+2i\eta-E_{ba}}\left(\frac{J^{\gamma}_{bc}J^{\beta}_{ca}f_{ac}}{\omega_1+i\eta-E_{ca}}-\frac{J^{\gamma}_{ca}J^{\beta}_{bc}f_{cb}}{\omega_1+i\eta-E_{bc}}\right)\Biggr],
\end{align}
where indices $a,b,c$ are spanned by the energy eigenvalues $E_a$ of the unperturbed BdG Hamiltonian $H_{\mathrm{BdG}}(\bm{k},\bm{A}=\bm{0})$ and the energy difference is $E_{ab} = E_{a} - E_{b}$. We introduced the Fermi-Dirac distribution function $f_{a}=(e^{\beta E_{a}}+1)^{-1}$ and defined $f_{ab} = f_{a} - f_{b}$. The infinitesimal positive parameter $\eta$ appears due to the adiabatic application of the external fields. The generalized electric current operators are defined by 
\begin{equation}
J^{\alpha_{1}\cdots\alpha_{n}}(\bm{k})=(-1)^{n}\left.\frac{\partial^{n}H_{\mathrm{BdG}}(\bm{k},\bm{A})}{\partial A^{\alpha_{1}}\cdots\partial A^{\alpha_{n}}}\right|_{\bm{A}=\bm{0}}.
\end{equation}
The first- and second-order ones ($n=1,2$) are called paramagnetic and diamagnetic current operators, respectively.

\subsection{Generalized Berry connection and paramagnetic current operator}
We begin with the discussion of the normal state.
Following discussions are useful to distinguish the optical responses of the normal and superconducting states, which are termed as normal and anomalous optical responses, respectively.

In the normal state, the Berry connection is related to the paramagnetic current operator $J^{\alpha}$.
With the Hellmann-Feynman theorem, the paramagnetic current operator is obtained as
\begin{equation}
\label{N_HF_rel}
J^{\alpha}_{ab}(\bm{k})=\frac{\partial}{\partial k_{\alpha}}E_{a}\delta_{ab}+iE_{ab}\xi^{\alpha}_{ab}(\bm{k}),
\end{equation}
with the Berry connection
 \begin{equation}
 \label{normal_Berry_connection}
\xi^{\alpha}_{ab}(\bm{k}) = i\left<u_{a}(\bm{k})\middle| \frac{\partial u_{b}(\bm{k})}{\partial k_{\alpha}}\right>,
\end{equation}
defined with the periodic part of the Bloch states $\{\ket{u_{a}(\bm{k})}\}$. This relation is essential for a remarkable simplification of formulas for transport and optical phenomena in the normal state~\cite{Watanabe2021,Watanabe2022,Sipe2000,Juan2020}.
The relation between the Berry connection and the paramagnetic current operator originates from the equivalence of differentiation with respect to the vector potential $\bm{A}$ and with respect to the momentum $-\bm{k}$, %The equivalence is 
which is justified due to the minimal coupling $\bm{p}\rightarrow\bm{p}-q\bm{A}$.

In contrast to the normal state, the Hellmann-Feynman relation in Eq.~\eqref{N_HF_rel} fails in the superconducting state. In the BdG formalism, particles with opposite charges, that is electron and hole, are treated on equal footing. %This treatment fails 
Thus, a naive treatment based on the minimal coupling $\bm{p}\rightarrow\bm{p}-q\bm{A}$ is not justified. However, we obtain similar relations between the Berry connection and paramagnetic current operator by introducing vector potential parametrization~\cite{Watanabe2022}. Bearing the minimal coupling in mind, we introduce the variational parameter $\bm{\lambda}$ by 
\begin{equation}
H_{\mathrm{BdG}}(\bm{k},\bm{A}=\bm{0})\rightarrow H_{\lambda}(\bm{k})=H_{\mathrm{BdG}}(\bm{k},\bm{\lambda}).
\end{equation}
Since the vector parameter $\bm{\lambda}$ plays the same role as the (spatially-uniform and time-independent) vector potential, we obtain the relation
\begin{equation}
J^{\alpha}_{ab}(\bm{k})=-\lim_{\bm{\lambda}\rightarrow\bm{0}}\Braket{a_{\bm{\lambda}}|\frac{\partial H_{\bm{\lambda}}}{\partial \lambda_{\alpha}}|b_{\bm{\lambda}}}.
\end{equation}
The Hellmann-Feynman relation for the vector potential parametrization is obtained from
\begin{equation}
\Braket{a_{\bm{\lambda}}|\frac{\partial H_{\bm{\lambda}}}{\partial \lambda_{\alpha}}|b_{\bm{\lambda}}}=\frac{\partial E_{a}(\bm{\lambda})}{\partial \lambda_{\alpha}}\delta_{ab}-i[\xi^{\lambda_{\alpha}},H_{\bm{\lambda}}]_{ab},
\end{equation}
where we defined the connection $\xi^{\lambda_{\alpha}}$ with the replacement $\partial_{k_{\alpha}} \rightarrow \partial_{\lambda_{\alpha}}$ in Eq.~\eqref{normal_Berry_connection}:
\begin{equation}
\xi^{\lambda_{\alpha}}_{ab}=i\left<a_{\bm{\lambda}}\middle| \frac{\partial b_{\lambda}}{\partial\lambda_{\alpha}}\right>.
\end{equation}
The relation between the paramagnetic current operator and Berry connection is as follows,
\begin{equation}
J^{\alpha}_{ab} = \lim_{\bm{\lambda}\rightarrow\bm{0}}\left[-\frac{\partial E_{a}(\bm{\lambda})}{\partial \lambda_{\alpha}}\delta_{ab}-iE_{ab}(\bm{\lambda})\xi^{\lambda_{\alpha}}_{ab}\right].
\end{equation}
This is the generalized Hellmann-Feynman relation in the superconducting state and plays an essential role in simplifying the optical conductivity formulas. %In addition, 
The linear optical conductivity and photocurrent response have been decomposed into the normal and anomalous contributions with using the vector potential  parametrization~\cite{Watanabe2022}.

\subsection{Anomalous photocurrent responses}
First, we discuss the photocurrent response given by the second-order nonlinear conductivity $\sigma^{\alpha;\beta\gamma}(0;\Omega,-\Omega)$ in this section.
In the gapful superconductors at low temperatures, the total photocurrent conductivity is decomposed into the two components;
\begin{equation}
\sigma = \sigma_{\mathrm{n}}+\sigma_{\mathrm{a}}.
\end{equation}
The former is a normal photocurrent which is a counterpart of the photocurrent in the normal state, while the latter is an anomalous photocurrent unique to the superconducting state.

The normal part consists of the four contributions
\begin{equation}
\sigma_{\mathrm{n}}=\sigma_{\mathrm{Einj}}+\sigma_{\mathrm{Minj}}+\sigma_{\mathrm{shift}}+\sigma_{\mathrm{gyro}},
\label{normal_photocurrent_total}
\end{equation}
which are termed electric injection current, magnetic injection current, shift current, and gyration current, respectively.
The formulas have been obtained as
\begin{align}
\label{eq:Einj_term}
\sigma^{\alpha;\beta\gamma}_{\mathrm{Einj}} &= -\frac{i\pi}{8\eta}\sum_{a\neq b}(J^{\alpha}_{aa}-J^{\alpha}_{bb})\Omega^{\lambda_{\beta}\lambda_{\gamma}}_{ba}F_{ab}, \\
\sigma^{\alpha;\beta\gamma}_{\mathrm{Minj}} &= \frac{\pi}{4\eta}\sum_{a\neq b}(J^{\alpha}_{aa}-J^{\alpha}_{bb})g^{\lambda_{\beta}\lambda_{\gamma}}_{ba}F_{ab}, \\
\sigma^{\alpha;\beta\gamma}_{\mathrm{shift}} &= -\frac{\pi}{4}\sum_{a\neq b}\Im\left[\left[D_{\lambda_{\alpha}}\xi^{\lambda_{\beta}}\right]_{ab}\xi^{\lambda_{\gamma}}_{ba} + \left[D_{\lambda_{\alpha}}\xi^{\lambda_{\gamma}}\right]_{ab}\xi^{\lambda_{\beta}}_{ba}\right]F_{ab}, \\
\sigma^{\alpha;\beta\gamma}_{\mathrm{gyro}} &= \frac{i\pi}{4}\sum_{a\neq b}\Re\left[\left[D_{\lambda_{\alpha}}\xi^{\lambda_{\beta}}\right]_{ab}\xi^{\lambda_{\gamma}}_{ba} - \left[D_{\lambda_{\alpha}}\xi^{\lambda_{\gamma}}\right]_{ab}\xi^{\lambda_{\beta}}_{ba}\right]F_{ab}, \label{eq:gyro_term}
\end{align}
where $F_{ab}=f_{ab}\delta(\Omega-E_{ba})$ means the Pauli exclusion principle at the optical transition. We also defined the geometric quantities such as the Berry curvature ($\Omega^{\lambda_{\alpha}\lambda_{\beta}}_{ab}$) and quantum metric ($g^{\lambda_{\alpha}\lambda_{\beta}}_{ab}$), and the covariant derivative ($D_{\lambda_{\alpha}}$). The covariant derivative is associated with the $U(N)$-Berry connection $\mathcal{A}^{\alpha}$ of N-fold energy-degenerate bands. When the pair potential is zero, we can easily reproduce the photocurrent response in the normal state by replacing $\bm{\lambda}$ with $\bm{k}$ in the normal photocurrent.
On the other hand, superconductivity may give a unique contribution to the normal photocurrent through the characteristic transition process, such as the type B in Fig.~\ref{fig:transition_image}. Indeed, we will see such photocurrent later.
%the normal photocurrent in the superconducting state is modified by new mechanisms such as Type B transition.

Because of the non-equivalence of $\bm{\lambda}$ and $\bm{k}$, the anomalous conductivity appears in the superconducting state. The anomalous part consists of two contributions
\begin{equation}
\sigma_{\mathrm{a}}=\sigma_{\mathrm{NRSF}}+\sigma_{\mathrm{CD}},
\end{equation}
which are termed nonreciprocal superfluid density term, and conductivity derivative term, respectively. The formulas are given by
\begin{align}
\label{NRSF}
\sigma^{\alpha;\beta\gamma}_{\mathrm{NRSF}} &= \lim_{\bm{\lambda}\rightarrow0}-\frac{1}{2\Omega^2}\partial_{\lambda_{\alpha}}\partial_{\lambda_{\beta}}\partial_{\lambda_{\gamma}}F_{\bm{\lambda}}, \\
\sigma^{\alpha;\beta\gamma}_{\mathrm{CD}} &= \lim_{\bm{\lambda}\rightarrow0}\frac{1}{4\Omega^{2}}\partial_{\lambda_{\alpha}}\left[\sum_{a\neq b}J^{\beta}_{ab}J^{\gamma}_{ba}f_{ab}\left(\frac{1}{\Omega-E_{ab}}+\frac{1}{E_{ab}}\right)\right]. \label{eq:CD_term}
\end{align}
In Eq.~\eqref{NRSF}, $F_{\bm{\lambda}}$ is the free energy of the BdG Hamiltonian.

%In Eqs.~\eqref{eq:Einj_term}-\eqref{eq:gyro_term} and \eqref{NRSF}-\eqref{eq:CD_term},
%the real and imaginary parts denote the photocurrent conductivity induced by linearly-polarized and circularly-polarized lights, respectively. Moreover, the contributions of photocurrent conductivity are classified in terms of the $\mathcal{T}$ and $\mathcal{PT}$ symmetry of the systems. For examples, the nonreciprocal superfluid density term is prohibited in the $\mathcal{T}$-symmetric system, while the conductivity derivative term disappears in the $\mathcal{PT}$-symmetric system.

A unique property of nonlinear optical responses of superconductors is the low-frequency divergence.
Indeed, the anomalous photocurrent shows divergent behaviors in the low-frequency regime. First, let us discuss the contribution from the nonreciprocal superfluid density. While the second-order derivative of the free energy is the superfluid density $\rho_{s}^{\alpha\beta}=\lim_{\bm{\lambda}\rightarrow\bm{0}}\partial_{\lambda_{\alpha}}\partial_{\lambda_{\beta}}F_{\bm{\lambda}}$, the nonreciprocal correction to the superfluid density $\rho_{s}^{\alpha\beta}$ is introduced as the nonreciprocal superfluid density $f^{\alpha\beta\gamma}=\lim_{\bm{\lambda}\rightarrow\bm{0}}\partial_{\lambda_{\alpha}}\partial_{\lambda_{\beta}}\partial_{\gamma}F_{\bm{\lambda}}$. An anomalous photocurrent in Eq.~\eqref{NRSF} is rewritten by 
\begin{equation}
\sigma^{\alpha;\beta\gamma}_{\mathrm{NRSF}} = -\frac{1}{2\Omega^{2}}f^{\alpha\beta\gamma},
\end{equation}
which shows the $\Omega^{-2}$ divergence at low frequencies.
Second, the conductivity derivative term Eq.~\eqref{eq:CD_term}
in the low-frequency limit ($\Omega\rightarrow0$) %the formula of conductivity derivative effect 
is reduced to the $\bm{\lambda}$-derivative of the total Berry curvature $\sum_{a}\epsilon_{\beta\gamma\delta}\Omega^{\lambda_{\delta}}_{a}f_{a}$,
\begin{eqnarray}
\sigma^{\alpha;\beta\gamma}_{\mathrm{CD}}\rightarrow\sigma^{\alpha;\beta\gamma}_{\mathrm{sCD}} &=&\lim_{\bm{\lambda}\rightarrow\bm{0}}\frac{i}{4\Omega}\epsilon_{\beta\gamma\delta}\partial_{\lambda_{\alpha}}\left(\sum_{a}\Omega^{\lambda_{\delta}}_{a}f_{a}\right), \\
&=& \frac{i}{4\Omega}\epsilon_{\beta\gamma\delta}B^{\alpha\delta}_{d}.
\label{eq:BCderivative}
\end{eqnarray}
Thus, the conductivity derivative term in the static limit ($\sigma^{\alpha;\beta\gamma}_{\mathrm{sCD}}$) is represented by \textit{Berry curvature derivative} $\hat{B}_{d}$ and 
shows the $\Omega^{-1}$ divergence.
%suppressed $O(\Omega^{0})$ terms. 
%$\sigma^{\alpha;\beta\gamma}_{\mathrm{NRSF}}$ and $\sigma^{\alpha;\beta\gamma}_{\mathrm{sCD}}$ is proportional to $\Omega^{-2}$ and $\Omega^{-1}$, respectively. 
Therefore, we expect that
the anomalous photocurrent responses enable the giant photo-electric conversion for the low-frequency light. 
The diverging behaviors are absent in normal metals and insulators and unique to superconductors.

In Eqs.~\eqref{eq:Einj_term}-\eqref{eq:gyro_term} and \eqref{NRSF}-\eqref{eq:BCderivative},
the real and imaginary parts denote the photocurrent conductivity induced by the linearly-polarized and circularly-polarized lights, respectively. Moreover, the contributions of photocurrent conductivity are classified in terms of the temporal $\mathcal{T}$ and $\mathcal{PT}$ symmetry of the systems~\cite{Watanabe2021,Watanabe2022,Sipe2000,Juan2020,Holder2020,Zhang2019}. For the anomalous photocurrent, the nonreciprocal superfluid density term is prohibited in the $\mathcal{T}$-symmetric system, while the Berry curvature derivative term disappears in the $\mathcal{PT}$-symmetric system.
%The anomalous conductivity is closely related to the temporal symmetry such as $\mathcal{T}$ and $\mathcal{PT}$ symmetries. 
This is because the nonreciprocal superfluid density $\hat{f}$ shows the odd-parity under the $\mathcal{T}$ symmetry operation, whereas the even-parity under the $\mathcal{PT}$ operation. In contrast, the Berry curvature derivative $\hat{B}_{d}$ is $\mathcal{T}$-even and $\mathcal{PT}$-odd. 

\subsection{Anomalous contribution to general nonlinear responses}
The low-frequency divergent behavior also emerges in other nonlinear responses such as the second harmonic generation. The anomalous parts of nonlinear responses have been obtained by the Green function method~\cite{Watanabe2022}. The low-frequency anomalous contribution reads
\begin{align}
&\sigma_\text{a}^{\alpha;\beta\gamma}(\omega_{1}+\omega_{2};\omega_{1},\omega_{2}) \notag \\
&=\sigma^{\alpha;\beta\gamma}_{\mathrm{NRSF}}(\omega_{1}+\omega_{2};\omega_{1},\omega_{2})+\sigma^{\alpha;\beta\gamma}_{\mathrm{CD}}(\omega_{1}+\omega_{2};\omega_{1},\omega_{2}),
\end{align}
where $O(\omega_{1}^{a}\omega_{2}^{b})$ terms are suppressed $(a+b\geq0)$.
The formulas of the nonreciprocal superfluid density and static conductivity derivative are obtain as
\begin{align}
\label{sNRSF}
&\sigma^{\alpha;\beta\gamma}_{\mathrm{NRSF}} = \frac{1}{2\omega_{1}\omega_{2}}f^{\alpha\beta\gamma}, \\
\label{sCD}
&\sigma^{\alpha;\beta\gamma}_{\mathrm{sCD}} \notag \\
&= -\frac{i}{4}\left(\frac{1}{\omega_{2}}\left(D^{\beta;\alpha\gamma}_{d}+\epsilon_{\alpha\gamma\delta}B^{\beta\delta}_{d}\right)+\frac{1}{\omega_{1}}\left(D^{\gamma;\alpha\beta}_{d}+\epsilon_{\alpha\beta\delta}B^{\gamma\delta}_{d}\right)\right).
\end{align}
In Eqs.~\eqref{sNRSF} and \eqref{sCD}, the nonreciprocal superfluid density $\hat{f}$, Berry curvature derivative $\hat{B}_{d}$, and Drude derivative $\hat{D}_{d}$ are defined in terms of Green function~\cite{Watanabe2022}. In contrast to the Berry curvature derivative, the Drude derivative is forbidden by the $\mathcal{T}$ symmetry though allowed in the $\mathcal{PT}$-symmetric systems.

Here, we consider the low-frequency anomalous contribution in a $\mathcal{T}$-symmetric two-dimensional superconductor, which we focus on later. The nonreciprocal superfluid density and Drude derivative are forbidden by the $\mathcal{T}$-symmetry, and thus, the anomalous contribution in the low-frequency limit reads
\begin{equation}
\sigma_\text{a}^{\alpha;\beta\gamma}(\omega_{1}+\omega_{2};\omega_{1},\omega_{2}) \to -\frac{i}{4}\left(\frac{1}{\omega_{2}}\epsilon_{\alpha\gamma\delta}B^{\beta\delta}_{d}+\frac{1}{\omega_{1}}\epsilon_{\alpha\beta\delta}B^{\gamma\delta}_{d}\right).
\end{equation}
The photocurrent conductivity $\sigma_{\mathrm{PC}}^{\alpha;\beta\gamma}$ and the second harmonic generation coefficient $\sigma_{\mathrm{SHG}}^{\alpha;\beta\gamma}$ are obtained as $\sigma_{\mathrm{PC}}^{\alpha;\beta\gamma}=\sigma^{\alpha;\beta\gamma}(0;\Omega, -\Omega)$ and $\sigma_{\mathrm{SHG}}^{\alpha;\beta\gamma}=\sigma^{\alpha;\beta\gamma}(2\Omega;\Omega;\Omega)$. The indices $\alpha,\beta,\gamma$ are $x$ or $y$ in the two-dimensional system. We obtain the following relations of nonlinear optical responses in the low-frequency regime
\begin{align}
\sigma^{x;xx}(\omega_{1}+\omega_{2};\omega_{1},\omega_{2}) &= \sigma^{y;yy}(\omega_{1}+\omega_{2};\omega_{1},\omega_{2})=0, \\
\sigma^{x;yy}_{\mathrm{PC}}&=\sigma^{y;xx}_{\mathrm{PC}}=0, \\
\sigma^{y;yx}_{\mathrm{SHG}} &= -\sigma^{y;yx}_{\mathrm{PC}}=-\frac{1}{2}\sigma^{x;yy}_{\mathrm{SHG}}, \\
\sigma^{x;xy}_{\mathrm{SHG}} &= -\sigma^{x;xy}_{\mathrm{PC}} = -\frac{1}{2}\sigma^{y;xx}_{\mathrm{SHG}}.
\end{align}
These relations are helpful in distinguishing the normal and anomalous contributions.

\section{Nonlinear optical responses in noncentrosymmetric superconductors}
\label{sec:method}

%\YYY{In the previous section, we have overviewed .....}

In the previous section, we have overviewed the formulation of second-order nonlinear responses in superconductors and the properties of anomalous contributions. In this section, we discuss the conditions where superconducting nonlinear responses appear.

\subsection{Model Hamiltonian}

In the following part, we investigate two-dimensional noncentrosymmetric superconductors with $\mathcal{T}$ symmetry. The model Hamiltonian consists of spinful fermions on the two-dimensional square lattice. By using the tight-binding approximation, the normal part $H_{\mathrm{N}}$ of the BdG Hamiltonian $H_{\mathrm{BdG}}$ is given by
\begin{equation}
H_{\mathrm{N}}(\bm{k}) = \xi_{\bm{k}} + \bm{g_{k}}\cdot\bm{\sigma},
\end{equation}
at each crystal momentum $\bm{k}$. The first term is kinetic energy measured from a chemical potential $\mu$, and the second term is an antisymmetric spin-orbit coupling. Pauli matrices $\bm{\sigma}$ denote the spin degree of freedom. Kinetic energy is assumed as 
\begin{equation}
\xi_{\bm{k}} = -2 t_{1}(\cos{k_{x}}+\cos{k_{y}}) + 4 t_{2}\cos{k_{x}}\cos{k_{y}} - \mu.
\end{equation}
The parity-mixed pair potential consists of the $s$-wave component $\psi_{\bm{k}}$ and the $p$-wave component $\bm{d_{k}}$.
Assuming the $\mathcal{T}$-symmetric parity-breaking superconductor, we have the pair potential in the $s$+$p$-wave form
\begin{equation}
\Delta_{\bm{k}} = (\psi_{\bm{k}}+\bm{d_{k}}\cdot\bm{\sigma})i\sigma_{y}.
\end{equation}
%at each crystal momentum $\bm{k}$. 

The symmetry of the system constrains the form of $\bm{g_{k}}$ and $\bm{d_{k}}$. We consider the two types of the forms, namely, \textit{s+p model} and \textit{extended s+p model}, which are allowed under the $m_{y}$ symmetry. Here, the $m_{y}$ symmetry denotes the mirror symmetry whose plane is perpendicular to the {\it y} axis. 
In the $s$+$p$ model, $\bm{g_{k}}$ and $\bm{d_{k}}$ are assumed as
\begin{align}
\bm{g_{k}} &= (\alpha_{1}\sin{k_{y}}, \alpha_{2}\sin{k_{x}}, \alpha_{3}\sin{k_{y}}), \\
\bm{d_{k}} &= (d_{1}\sin{k_{y}}, d_{2}\sin{k_{x}}, d_{3}\sin{k_{y}}),
\end{align}
while the extended $s$+$p$ model defines them as 
\begin{align}
\bm{g_{k}} &= (\alpha_{1}\sin{2k_{y}}, \alpha_{2}\sin{k_{x}}, \alpha_{3}\sin{k_{y}}), \\
\bm{d_{k}} &= (d_{1}\sin{2k_{y}}, d_{2}\sin{k_{x}}, d_{3}\sin{k_{y}}). 
\end{align}
Although these models are similar to each other, there is an important difference: in the normal state the second-order optical responses are absent in the $s$+$p$ model while present in the extended $s$+$p$ model. 
%The important difference between these models is whether the second-order nonlinear response of the normal state can be nonzero. 
As we show below, the second-order conductivity in the normal state $\sigma_{N}$ disappears in the $s$+$p$ model for arbitrary $\alpha_{1},\cdots,\alpha_{3}$, but it can be finite in the extended $s$+$p$ model. Therefore, the second-order optical responses of the $s$+$p$ model essentially originate from the superconductivity.

In the numerical calculations, the parameters are set as $\mu=-0.8$, $t_{1}=1$, and $t_{2}=0.2$. For a quantitative estimation, we set $t_{1}=1~\mathrm{eV}$ and calculate the response coefficients in the SI unit. Thus, the numerical results of the nonlinear conductivity are given in the unit $\mathrm{A\cdot V^{-2}}$. Numerical calculations are performed on the $N^{2}$-discretized Brillouin zone ($N=3000$). For numerical convergence, we introduce a phenomenological scattering rate $\gamma=10^{-4}$ and a finite temperature $T=10^{-4}$ for the Fermi-Dirac distribution function.

\subsection{Condition of finite nonlinear optical response}\label{subsec:condition_Einj}
In this subsection, we show a conjecture on the microscopic condition for the nonzero photocurrent conductivity of the two-dimensional $\mathcal{T}$-symmetric systems. Conditions beyond symmetry arguments are obtained as follows.

First, we discuss the condition in the normal state. In the zero-temperature limit $T\rightarrow0$, the electric injection current of the normal state is written as 
\begin{align}
\label{normal_inj}
\sigma_{\mathrm{Einj}}^{\alpha;\beta\gamma} = \frac{-i\pi}{4\gamma}&\sum_{\bm{k}} (\partial_{\alpha}\bm{g_{k}}\cdot\hat{\bm{g}}_{\bm{k}})\frac{[(\partial_{\beta}\bm{g_{k}}\times\partial_{\gamma}\bm{g_{k}})\cdot\hat{\bm{g}}_{\bm{k}}]}{g_{\bm{k}^{2}}} \notag \\
&\times\Theta(g_{\bm{k}}-\xi_{\bm{k}})\delta(\Omega-2 g_{\bm{k}}),
\end{align}
with the scattering rate $\gamma$. The step function $\Theta$ appeared due to the zero temperature limit. Here, $\hat{\bm{g}}_{\bm{k}}$ and $\partial_{\alpha}$ means an unit vector defined as $\hat{\bm{g}}_{\bm{k}}=\bm{g_{k}}/|\bm{g_{k}}|$ and $k_{\alpha}$ derivative $\partial_{k_{\alpha}}$, respectively. Eq.~\eqref{normal_inj} implies that the condition
%$[
\begin{align}
    (\partial_{\beta}\bm{g_{k}}\times\partial_{\gamma}\bm{g_{k}})\cdot\hat{\bm{g}}_{\bm{k}}\neq0,
\end{align}
%$ 
is essential for the nonzero injection current. The factor $[(\partial_{\beta}\bm{g_{k}}\times\partial_{\gamma}\bm{g_{k}})\cdot\hat{\bm{g}}_{\bm{k}}]$ is derived from the Berry curvature, which we call $\textit{normal Berry curvature factor}$. The normal Berry curvature factor is also essential for the total photocurrent conductivity.

When the normal Berry curvature factor is zero, the total photocurrent conductivity $\sigma^{\alpha;\beta\gamma}_{\mathrm{N}}$ is given by
\begin{equation}
\sigma^{\alpha;\beta\gamma}_{\mathrm{N}}=-\sum_{\bm{k}}\frac{\pi}{4g_{\bm{k}}^{2}}[(\partial_{\alpha\beta}\bm{g_{k}}\times\partial_{\gamma}\bm{g_{k}})\cdot\hat{\bm{g}}_{\bm{k}}]\Theta(g_{\bm{k}}-\xi_{\bm{k}})\delta(\Omega-2 g_{\bm{k}}),
\end{equation}
where $\partial_{\alpha\beta}= \partial_{k_\alpha} \partial_{k_\beta}$.
According to Eq.~\eqref{normal_photocurrent_total}, this contribution is given by the shift current. 
In the cases of the $s$+$p$ and extended $s$+$p$ models, the factor $[(\partial_{\alpha\beta}\bm{g_{k}}\times\partial_{\gamma}\bm{g_{k}})\cdot\hat{\bm{g}}_{\bm{k}}]$ vanishes under the condition that the normal Berry curvature factor is zero. Thus, the photocurrent conductivity %of the $s$+$p$ and extended $s$+$p$ model 
is nonzero only when the normal Berry curvature factor is nonzero. The same rule applies to other simple models as well. Thus, the normal Berry curvature factor is important for the photocurrent conductivity in the normal state.

Next, we discuss the condition for the nonzero photocurrent conductivity in the superconducting state. Again we consider the electric injection current, which is rewritten as
\begin{align}
\sigma^{\alpha;\beta\gamma}_{\mathrm{Einj}}=\frac{\pi}{4\eta}\sum_{\bm{k}}&\sum_{a,b}\left[J^{\alpha}_{aa}(\bm{k})-J^{\alpha}_{bb}(\bm{k})\right]\frac{J^{\beta}_{ba}(\bm{k})J^{\gamma}_{ab}(\bm{k})}{{E_{ba}}^{2}} \notag \\
&\times f_{ab}\delta\left(\Omega-E_{ba}\right).
\end{align}
%in the superconducting state.
Here, we approximate eigenstates and eigenvalues of the Hamiltonian. We decompose the normal part of BdG Hamiltonian 
\begin{equation}
H_{\mathrm{N}}(\bm{k})=\xi_{\bm{k}}+\bm{g}_{\parallel\bm{k}}+\bm{g}_{\perp\bm{k}}=\tilde{H}_{N}(\bm{k})+\bm{g}_{\perp\bm{k}},
\end{equation}
where the decomposition of $\bm{g_{k}}$ is defined by
\begin{equation}
\bm{g_{k}}=\bm{g}_{\parallel\bm{k}}+\bm{g}_{\perp\bm{k}}, \qquad \bm{g}_{\parallel\bm{k}}\times \bm{d_{k}}=0, \quad \bm{g}_{\perp\bm{k}}\cdot \bm{d_{k}}=0.
\end{equation}
The approximate BdG Hamiltonian $\tilde{H}_{\mathrm{BdG}}$ is defined by
\begin{equation}
\tilde{H}_{\mathrm{BdG}}(\bm{k}) = 
\begin{pmatrix}
\tilde{H}_{N}(\bm{k}) & \Delta_{\bm{k}} \\
\Delta^{\dagger}_{\bm{k}} & -\left[\tilde{H}_{N}(-\bm{k})\right]^{T} \\
\end{pmatrix},
\end{equation}
where the pair potential manifests only the intraband components in the band representation for the Hamiltonian $\tilde{H}_{N}(\bm{k})$. 
Using the eigenvalues $\tilde{E}_{a}$ and eigenvectors $\ket{\tilde{a}_{\bm{\lambda}}}$ of $\tilde{H}_{\mathrm{BdG}}$, we approximate the velocity operator $J^{\alpha}_{ab}(\bm{k})$ by
\begin{equation}
\tilde{J}^{\alpha}_{ab}(\bm{k})=-\lim_{\bm{\lambda}\rightarrow\bm{0}}\Braket{\tilde{a}_{\bm{\lambda}}|\frac{\partial H_{\bm{\lambda}}}{\partial \lambda_{\alpha}}|\tilde{b}_{\bm{\lambda}}}.
\end{equation}
Now an approximate formula for the electric injection current, %$\sigma^{\alpha;\beta\gamma}_{\mathrm{Einj}}$, %by using the eigenvalues $\tilde{E}_{a}$ and eigenvectors $\ket{\tilde{a}_{\bm{\lambda}}}$.
\begin{align}
\sigma^{\alpha;\beta\gamma}_{\mathrm{Einj}}\simeq\frac{\pi}{4\eta}\sum_{\bm{k}}&\sum_{a,b}\left[\tilde{J}^{\alpha}_{aa}(\bm{k})-\tilde{J}^{\alpha}_{bb}(\bm{k})\right]\frac{\tilde{J}^{\beta}_{ba}(\bm{k})\tilde{J}^{\gamma}_{ab}(\bm{k})}{{\tilde{E}_{ba}}^{2}} \notag \\
&\times\tilde{f}_{ab}\delta\left(\Omega-\tilde{E}_{ba}\right),
\end{align}
can be analytically calculated,
where $\tilde{f}_{ab}$ is the difference of Fermi-Dirac distributions between the eigenvalues $\tilde{E}_{a}$ and $\tilde{E}_{b}$.
In this approximation, we obtain the analytic formula for the electric injection current $\tilde{\sigma}^{\alpha;\beta\gamma}_{\mathrm{Einj}}$ as
\begin{align}
\label{appro_Einj}
\tilde{\sigma}^{\alpha;\beta\gamma}_{\mathrm{Einj}} = -\frac{i\pi}{4\eta}\sum_{\bm{k}}&\frac{\left[F_{+}(-\psi_{\bm{k}}+d_{\bm{k}})+F_{-}(\psi_{\bm{k}}+d_{\bm{k}})\right]^{2}}{E_{+}E_{-}F_{+}F_{-}(E_{+}+E_{-})^{2}}(\partial_{\alpha}\bm{g_{k}}\cdot\hat{\bm{d_{k}}}) \notag \\
&\times \left[(\partial_{\beta}\bm{g_{k}}\times\partial_{\gamma}\bm{g_{k}})\cdot\hat{\bm{d_{k}}}\right]\delta\left(\Omega-(E_{+}+E_{-})\right),
\end{align}
where we defined $E_{\pm}$ and $F_{\pm}$ as
\begin{align}
E_{\pm}&=\sqrt{(\xi_{\bm{k}}\pm\bm{g_{k}}\cdot\bm{\hat{d}_{k}})^{2}+(\psi_{\bm{k}}\pm d_{\bm{k}})^{2}}, \\
F_{\pm}&=E_{\pm}+\xi_{\bm{k}}\pm \bm{g_{k}}\cdot\bm{\hat{d}_{k}}.
\end{align}
According to Eq.~\eqref{appro_Einj}, the approximate electric injection current vanishes when $\left[(\partial_{\beta}\bm{g_{k}}\times\partial_{\gamma}\bm{g_{k}})\cdot\hat{\bm{d_{k}}}\right]$ is zero. 
This analytical result implies that the factor $\left[(\partial_{\beta}\bm{g_{k}}\times\partial_{\gamma}\bm{g_{k}})\cdot\hat{\bm{d_{k}}}\right]$, which we call $\textit{superconducting Berry curvature factor}$ (SC Berry curvature factor), is essential for the electric injection current. Similarly to the normal state responses, the SC Berry curvature factor is also essential for other nonlinear optical responses. In Sec.~\ref{sec:result}, the relation between the SC Berry curvature factor and nonlinear optical conductivities such as the total photocurrent and second harmonic generations is numerically elucidated.

\subsection{Intraband pairing and interband pairing}

Before showing the numerical results, we discuss the mechanism and condition of the superconducting nonlinear optical response based on the band representation of pair potential.
A pair potential $\Delta_{\bm{k}}$ consists of intraband and interband pairing, which can be captured by the superconducting fitness~\cite{Ramires2016,Ramires2018}. The intraband components of the pair potential are $\psi_{\bm{k}}$ and $\bm{d}_{\parallel\bm{k}}$, while the interband one is $\bm{d}_{\perp\bm{k}}$ with the decomposition of $\bm{d}_{\bm{k}}$ defined as 
\begin{equation}
\bm{d_{k}}=\bm{d}_{\perp\bm{k}}+\bm{d}_{\parallel\bm{k}}, \quad \bm{d}_{\parallel\bm{k}}\times \bm{g_{k}}=0, \quad \bm{d}_{\perp\bm{k}}\cdot\bm{g_{k}}=0.
\end{equation}
When the pair potential consists of only intraband components, the electric injection current is exactly obtained as
\begin{align}
\label{intra_Einj}
\sigma^{\alpha;\beta\gamma}_{\mathrm{Einj}} = -\frac{i\pi}{4\eta}\sum_{\bm{k}}&\frac{\left[F_{+}(-\psi_{\bm{k}}+d_{\bm{k}})+F_{-}(\psi_{\bm{k}}+d_{\bm{k}})\right]^{2}}{E_{+}E_{-}F_{+}F_{-}(E_{+}+E_{-})^{2}}(\partial_{\alpha}\bm{g_{k}}\cdot\hat{\bm{g}}_{\bm{k}}) \notag \\
&\times \left[(\partial_{\beta}\bm{g_{k}}\times\partial_{\gamma}\bm{g_{k}})\cdot\hat{\bm{g}}_{\bm{k}}\right]\delta\left(\Omega-(E_{+}+E_{-})\right),
\end{align}
where we defined $E_{\pm}$ and $F_{\pm}$ as
\begin{align}
E_{\pm}&=\sqrt{(\xi_{\bm{k}}\pm g_{\bm{k}})^{2}+(\psi_{\bm{k}}\pm \bm{d_{k}}\cdot\hat{\bm{g}}_{\bm{k}})^{2}}, \\
F_{\pm}&=E_{\pm}+\xi_{\bm{k}}\pm g_{\bm{k}}.
\end{align}
On the other hand, when the pair potential consists of only interband components, the electric injection current is given by
\begin{align}
\label{inter_Einj}
\sigma^{\alpha;\beta\gamma}_{\mathrm{Einj}}&=\frac{-i\pi}{4\eta}\sum_{\bm{k}}\frac{\xi_{\bm{k}}(\partial_{\alpha}\bm{g_{k}}\cdot\hat{\bm{g}}_{\bm{k}})}{u_{\bm{k}}g_{\bm{k}}^{2}}\left(1-\frac{d_{\bm{k}}^{2}}{u_{\bm{k}}(u_{\bm{k}}+\xi_{\bm{k}})}\right) \notag \\
\times&\left[\left(\partial_{\beta}\bm{g_{k}}\times\partial_{\gamma}\bm{g_{k}}\right)\cdot\hat{\bm{g}}_{\bm{k}}\right]\Theta\left(g_{\bm{k}}-\sqrt{\xi_{\bm{k}}^{2} + d_{\bm{k}}^{2}}\right) %\notag \\ &\quad\times
\delta\left(\Omega-2 g_{\bm{k}}\right),
\end{align}
%where we defined $u_{\bm{k}}$ as
with 
\begin{equation}
u_{\bm{k}}=\sqrt{\xi_{\bm{k}}^{2}+d_{\bm{k}}^{2}}.
\end{equation}

From Eqs.~\eqref{normal_inj},  \eqref{intra_Einj}, and \eqref{inter_Einj}, we find similarity of the electric injection current between the normal state and the purely intraband or interband pairing state. 
When either an intraband component or an interband component of pair potential is absent, the normal Berry curvature factor must be nonzero to obtain a finite injection current. This condition has also been shown for the normal state. Furthermore, 
the injection current contributions in Eqs.~\eqref{intra_Einj} and \eqref{inter_Einj} arise from the quasiparticle excitation labeled by the off-diagonal elements in the Bogoliubov spectrum.
When the pair potential is sufficiently small, the resonant frequency is $\Omega \sim 2g_{\bm{k}}$ as in Eq.~\eqref{normal_inj} for the normal state. These similarities imply that the electric injection current in the purely intraband or interband pairing state (Eq.~\eqref{intra_Einj} and \eqref{inter_Einj}) are similar to the contribution existing in the normal state. Therefore, we expect that coexistence of the intraband and interband pairing is essential for the nonlinear optical responses unique to the superconducting state. This expectation is justified in Sec.~\ref{sec:result}, where the relation between the pair potential and nonlinear optical conductivities such as the total photocurrent and second harmonic generations is investigated numerically and comprehensively.

\section{Numerical Results}
\label{sec:result}
In this section, we demonstrate various nonlinear optical conductivities in noncentrosymmetric superconductors with numerical calculations. We consider the $s$+$p$ model and the extended $s$+$p$ model, assuming the $\mathcal{T}$ symmetry and $m_{y}$ symmetry of the system. Contributions unique to superconductors appear in the numerical results of the photocurrent generation and second harmonic generation. We elucidate the relation between the normal and SC Berry curvature factors and the nonlinear optical responses. We will see that the presence of finite Berry curvature factors is an essential condition for the nonlinear responses. Furthermore, we clarify the essential role of intraband and interband pairing in the nonlinear optical responses. 

\subsection{General property of $s$+$p$ model and extended $s$+$p$ model}

\begin{figure*}[tbp]
 \includegraphics[width=\linewidth]{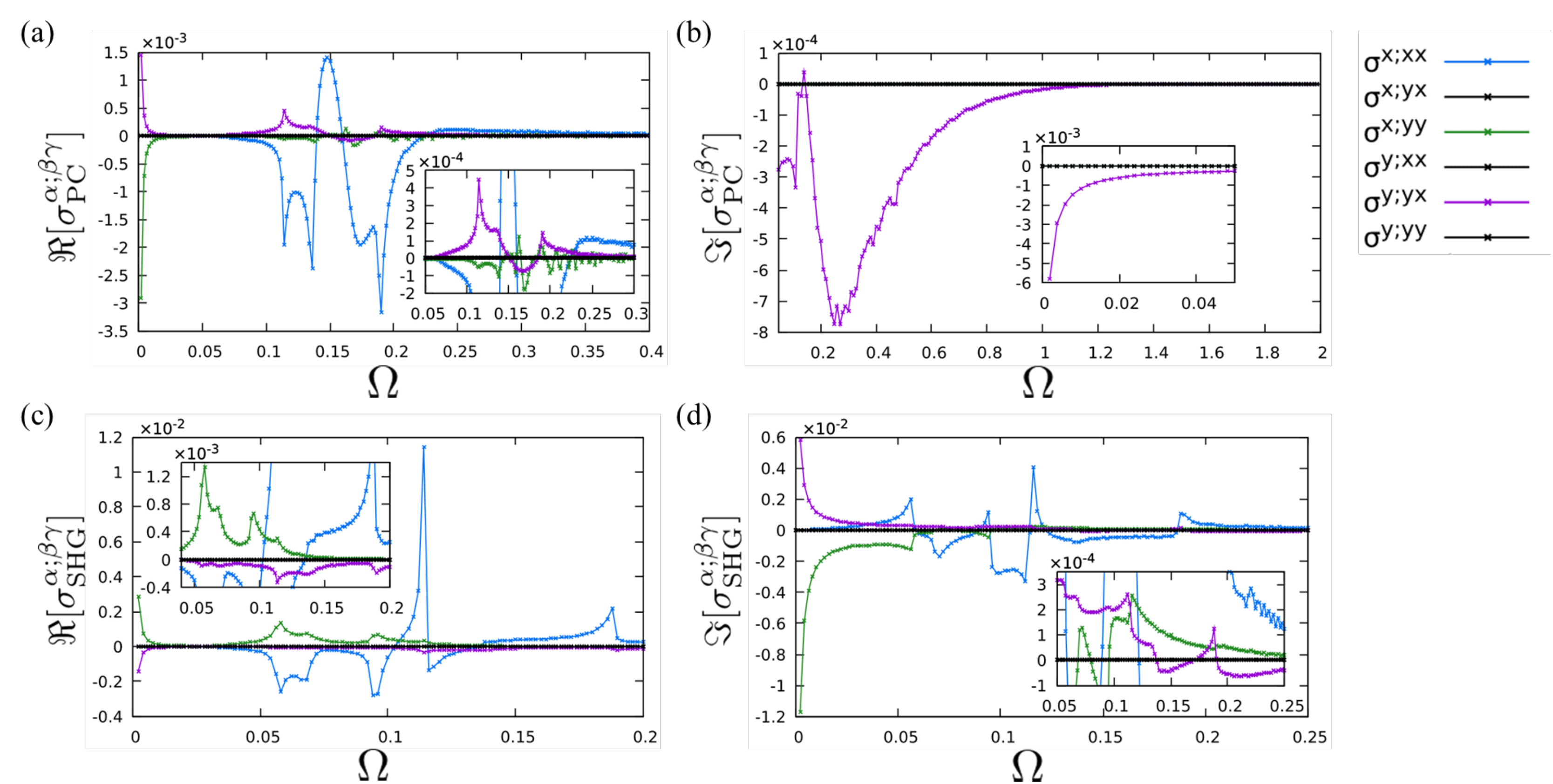}
 \caption{Frequency dependence of the photocurrent generation and second harmonic generation in the $s$+$p$ model. (a) Real part $\Re[\sigma^{\alpha;\beta\gamma}_{\mathrm{PC}}]$ and (b) imaginary part $\Im[\sigma^{\alpha;\beta\gamma}_{\mathrm{PC}}]$ of the photocurrent generation. (c) Real part $\Re[\sigma^{\alpha;\beta\gamma}_{\mathrm{SHG}}]$ and (d) imaginary part $\Im[\sigma^{\alpha;\beta\gamma}_{\mathrm{SHG}}]$ of the second harmonic generation. We take the parameter set of No.1 in Table~\ref{tab:s_p_model_para}, $\alpha_{1}=0.3$, $\alpha_{2}=0.8$, $\alpha_{3}=0.1$, $\psi_{0}=0.07$, $d_{1}$=0.04, $d_{2}$=0.02, and $d_{3}$=0.05. Note that $\sigma^{x;xy}=\sigma^{y;xx}=\sigma^{y;yy}=0$ as those components are prohibited by the $m_y$ mirror symmetry.}
 \label{fig:gene_pro}
\end{figure*}

Before showing the main results of this section, %investigating the roles of the pair potential and the normal and SC Berry curvature factors, 
we discuss some general properties of the nonlinear responses in the models. %the $s$+$p$ model and extended $s$+$p$ model. 
There are some constraints for the photocurrent conductivity $\sigma^{\alpha;\beta\gamma}_{\mathrm{PC}}$ and second harmonic generation $\sigma^{\alpha;\beta\gamma}_{\mathrm{SHG}}$. The first constraint %is given by 
\begin{equation}
\sigma^{\alpha;\beta\gamma}_{\mathrm{PC}}=\left(\sigma^{\alpha;\gamma\beta}_{\mathrm{PC}}\right)^{*}, \qquad \sigma^{\alpha;\beta\gamma}_{\mathrm{SHG}}=\sigma^{\alpha;\gamma\beta}_{\mathrm{SHG}},
\end{equation}
has to be satisfied. 
% because the electric current $\left<\mathcal{J}^{\alpha}(t)\right>$ is real in Eq.~\eqref{eq:2nd_nonlinear_res}.
This constraint requires ${\Im}[\sigma^{\alpha;\beta\beta}_{\mathrm{PC}}]=0$, which is consistent with the numerical results. The second constraint is owing to the crystal symmetry of the system.
The $s$+$p$ model and extended $s$+$p$ model are characterized by the $m_{y}$ mirror symmetry. This symmetry allows the nonlinear conductivity, $\sigma^{x;xx}$, $\sigma^{x;yy}$, and $\sigma^{y;yx}$ in the two-dimensional system.

Figure~\ref{fig:gene_pro} plots the photocurrent conductivity $\sigma_{\mathrm{PC}}^{\alpha;\beta\gamma}=\sigma^{\alpha;\beta\gamma}(0;\Omega, -\Omega)$ and second harmonic generation $\sigma_{\mathrm{SHG}}^{\alpha;\beta\gamma}=\sigma^{\alpha;\beta\gamma}(2\Omega;\Omega,\Omega)$ in the $s$+$p$ model, which satisfy the above constraints.
The divergent nonlinear responses are observed in the low-frequency regime. The static conductivity derivative $\sigma^{\alpha;\beta\gamma}_{\mathrm{sCD}}$, which is allowed by the $\mathcal{T}$ symmetry, causes the anomalous divergent behaviors in the imaginary part of $\sigma_{\mathrm{PC}}$ and $\sigma_{\mathrm{SHG}}$. Although the real part also shows a diverging behavior, this contribution is artificial and comes from the phenomenological treatment of the scattering rate~\cite{Watanabe2022}.
%it is dependent significantly on the scattering rate. 
%It indicates that the divergence of the real part is the extrinsic contribution of the scattering effect.

In the following subsections, we discuss the relationship between the nonlinear optical response and the presence of intraband and interband pairing. In the $s$+$p$ and extended $s$+$p$ models, the photocurrent and second harmonic generation unique to the superconducting state can appear when the pair potential includes both intraband and interband pairing components. Comparison to the joint density of states indicates that $E\leftrightarrow-E$ transition (type B in Fig.~\ref{fig:transition_image}) dominantly contributes to the superconducting nonlinear optical responses.

\subsection{$s$+$p$ model}
\label{subsec:s+p_model}

\begin{figure*}[tbp]
 \includegraphics[width=\linewidth]{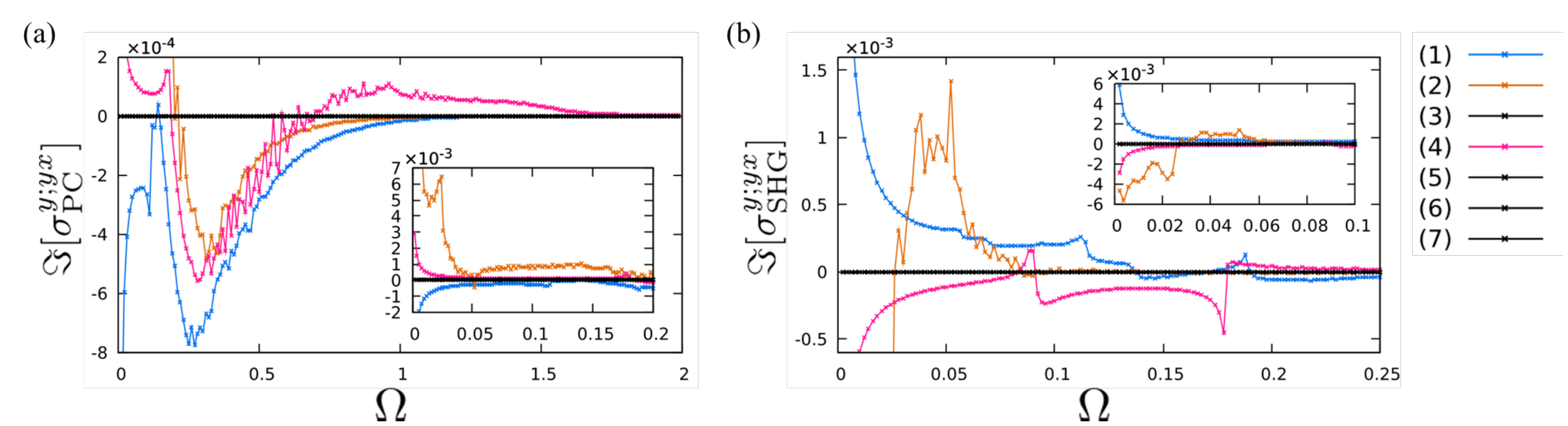}
 \caption{Frequency dependence of the imaginary part of (a) photocurrent conductivity $\sigma^{y;yx}_{\mathrm{PC}}$ and (b) second harmonic generation $\sigma^{y;yx}_{\mathrm{SHG}}$ in the $s$+$p$ model with varying the parameters $(\alpha_{1}, \alpha_2,\alpha_{3})$, $\psi_{0}$, and $(d_{1}, d_2, d_{3})$. Table~\ref{tab:s_p_model_para} shows the parameter sets of the cases (1)-(7). Presence or absence of the nonlinear optical responses are summarized in Table~\ref{tab:s_p_model_para}.}
 \label{fig:s_p_model_para}
\end{figure*}

\begin{table*}[tbp]
\centering
\begin{tabular}{|c|c|c|c|c|c|c|c|c|c|}
\hline
No. &  $(\alpha_{1},\alpha_{2},\alpha_{3})$ & $\psi_{0}$ & $(d_{1}, d_{2}, d_{3})$ & $\bm{d}_{\bm{k}}\cdot\bm{g}_{\bm{k}}$ & $\bm{d}_{\bm{k}}\times\bm{g}_{\bm{k}}$& $(\partial_{x}\bm{g_{\bm{k}}}\times\partial_{y}\bm{g_{\bm{k}}})\cdot\hat{\bm{d}}_{\bm{k}}$ & $\Delta_{\mathrm{intra}}$ & $\Delta_{\mathrm{inter}}$ & response \\
 \hline
1 & (0.3, 0.8, 0.1) & 0.07 & (0.04, 0.02, 0.05) & nonzero & nonzero & nonzero & nonzero & nonzero & nonzero \\
\hline
2 & (0.3, 0.8, 0.1) & 0 &  (0.04, 0.02, 0.05) & nonzero & nonzero & nonzero & nonzero & nonzero & nonzero \\
 \hline
3 & (0.2, 0.9, 0.1) & 0.01 & (0.06, 0.05, 0.03) & nonzero & nonzero & zero & nonzero & nonzero & zero \\
 \hline
4 & (0.7, 0.2, 0.3) & 0.09 & (0.03, 0, -0.07) & zero & nonzero & nonzero & nonzero & nonzero & nonzero \\
\hline
5 & (0.7, 0.2, 0.3) & 0 & (0.03, 0, -0.07) & zero & nonzero & nonzero & zero & nonzero & zero \\
\hline
6 & (0.4, 0.1, 0.6) & 0.09 & (0, 0, 0) & zero & zero & undefined & nonzero & zero & zero \\
\hline
7 & (0.3, 0.2, 0.6) & 0 & (0, 0, 0) & zero & zero & undefined & zero & zero & zero \\
\hline
\end{tabular}
\caption{The parameter set $(\alpha_{1}, \alpha_2, \alpha_{3})$, $\psi_{0}$, and $(d_{1}, d_2, d_{3})$ adopted in Fig.~\ref{fig:s_p_model_para} and existence or non-existence of nonlinear optical responses (first right column). When $\bm{d}_{\bm{k}}\cdot\bm{g}_{\bm{k}}$ or $\psi_0$ is nonzero, %$\bm{d}_{\bm{k}}$ 
the pair potential has an intraband component (third right column). When $\bm{d}_{\bm{k}} \times \bm{g}_{\bm{k}} \ne 0$, the interband component is finite (second right column). The fourth right column shows whether the SC Berry curvature factor is zero or nonzero.}
\label{tab:s_p_model_para}
\end{table*}

Here, numerical analysis of the photocurrent conductivity %$\sigma^{\alpha;\beta\gamma}(0;\Omega, -\Omega)$ 
and second harmonic generation %$\sigma^{\alpha;\beta\gamma}(2\Omega;\Omega,\Omega)$ 
in the $s$+$p$ model is presented in details.
We compare the numerical results with varying conditions about normal and SC Berry curvature factors and spin-singlet and spin-triplet components of pair potential.
In the $s$+$p$ model, the normal and SC Berry curvature factors are given by
\begin{align}
\label{eq:normal_Berry_curvature_sp}
\partial_{x}\bm{g_{k}}\times\partial_{y}\bm{g_{k}}\cdot\hat{\bm{g_{k}}} &= 0, \\
\label{eq:SC_Berry_curvature_sp}
\partial_{x}\bm{g_{k}}\times\partial_{y}\bm{g_{k}}\cdot\hat{\bm{d_{k}}} &\propto \alpha_{2}(d_{1}\alpha_{3}-d_{3}\alpha_{1})\sin k_{y}\cos k_{x} \cos k_{y},
\end{align}
respectively. As shown in Eq.~\eqref{eq:normal_Berry_curvature_sp}, the normal Berry curvature factor vanishes in the $s$+$p$ model, and therefore, the nonlinear optical responses are absent in the normal state (No.7 case in Table~\ref{tab:s_p_model_para}).
Furthermore, the SC Berry curvature factor also disappears 
%is zero 
when the pair potential consists of only the intraband pairing component, that is, %because intraband components of 
$\bm{d_{k}}$ is parallel to $\bm{g_{k}}$. 

Figure \ref{fig:s_p_model_para} plots the photocurrent conductivity $\sigma^{y;yx}_{\mathrm{PC}}$ and second harmonic generation $\sigma^{y;yx}_{\mathrm{SHG}}$ for various parameter sets of spin-orbit coupling, $(\alpha_{1}, \alpha_2, \alpha_{3})$, and pair potential, $\psi_{0}$ and $(d_{1}, d_2, d_{3})$.
The parameter sets labeled as No.1-No.7 and whether the second-order nonlinear response is zero or nonzero are summarized in Table~\ref{tab:s_p_model_para}. In the No.1, 2, and 4 cases, the nonlinear optical responses appear and show diverging behavior in the low-frequency region. On the other hand, the photocurrent and second harmonic generation disappear in the No.3 case, where the SC Berry curvature factor is zero. Thus, the nonzero SC Berry curvature factor is essential for the second-order nonlinear responses in the $s$+$p$ model. This condition has been analytically derived for the injection current in Sec.~\ref{subsec:condition_Einj}. The numerical results imply that the condition applies to the total nonlinear conductivity as well. 

Next, we discuss the roles of intraband pairing and interband pairing.
The No.1, 2, and 4 cases, in which the nonlinear responses are nonzero, have both the intraband and interband pairing components of the pair potential. On the other hand, the second-order nonlinear responses disappear in the No.5 case, though the difference from the No.4 case is only the absence of the spin-singlet pairing component $\psi_0$. 
%$\bm{d_{k}}$ is parallel to $\bm{g_{k}}$. 
Note that the spin-singlet pairing component is not necessarily required for the nonlinear optical responses to appear as they indeed appear in the No.2 case. 
%which also does not has a spin-singlet pairing. 
%the nonlinear response appears unlike in the No.5 case. The difference between No.2 and No.5 is whether the intraband pairing is nonzero. 
The essential difference of the No.5 case from the No.2 and 4 cases is the absence of the intraband pairing components in the pair potential, namely, $\bm{d}_{\parallel \bm{k}}$ and $\psi_0$. Thus, it is indicated that the intraband pairing is required for the second-order nonlinear responses. 
The interband pairing is also needed because the SC Berry curvature factor is zero when the pair potential consists of only the intraband pairing components. Therefore, both intraband and interband pairing are necessary for the photocurrent and second harmonic generations unique to superconductors. 
This is one of the main conclusions of this paper.
%in the single-band models for noncentrosymmetric superconductors.

\begin{figure}[tbp]
 \includegraphics[width=\linewidth]{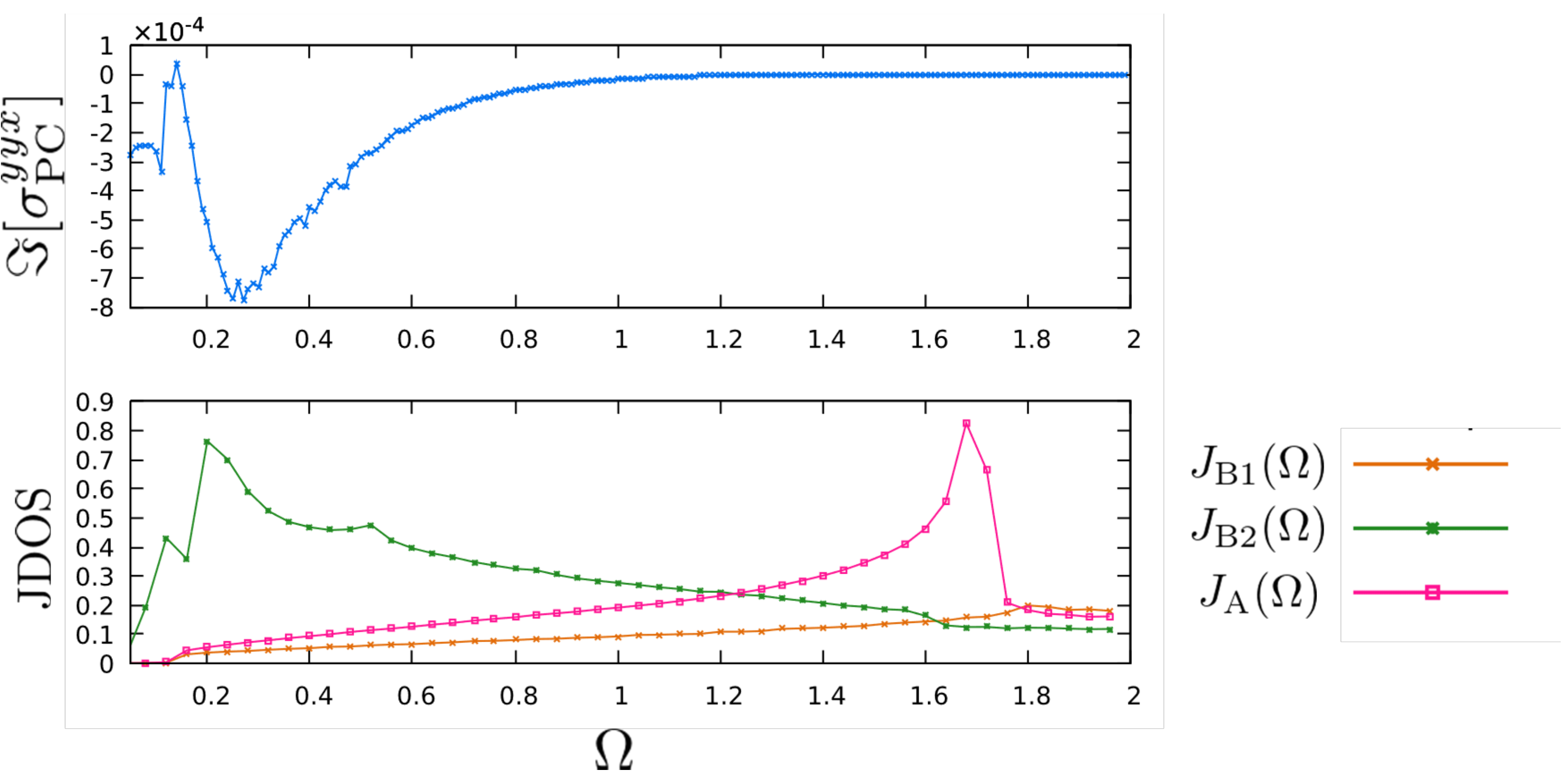}
 \caption{(Upper panel) frequency dependence of the imaginary part of the photocurrent conductivity $\sigma^{y;yx}_{\mathrm{PC}}$ in the $s$+$p$ model. Parameters are the same as the No.1 case in Table~\ref{tab:s_p_model_para}.
 (Lower panel) joint density of states for each transition process.  $J_{\mathrm{B1}}(\Omega)$, $J_{\mathrm{B2}}(\Omega)$, and $J_{\mathrm{A}}(\Omega)$ are the joint density of states for $E_{1}\leftrightarrow -E_{1}$, $E_{2}\leftrightarrow -E_{2}$, and $E_{i}\leftrightarrow -E_{j} \quad (i\neq j)$ transition, respectively. $E_{1}$ and $E_{2}$ are positive eigenenergies of Bogoliubov quasiparticles ($E_{1}\geq E_{2} \geq 0$).}
 \label{fig:s_p_model_DOS_Im_tot}
\end{figure}

Figure~\ref{fig:s_p_model_DOS_Im_tot} shows the frequency dependence of the imaginary part of photocurrent conductivity $\sigma^{y;yx}_{\mathrm{PC}}$ and the joint density of states. The joint density of states $J_{\mathrm{B1}}(\Omega)$, $J_{\mathrm{B2}}(\Omega)$, and $J_{\mathrm{A}}(\Omega)$ are defined as $J_{\mathrm{B1}}(\Omega)=\sum_{\bm{k}}\delta(\Omega-2E_{1\bm{k}})$, $J_{\mathrm{B2}}(\Omega)=\sum_{\bm{k}}\delta(\Omega-2E_{2\bm{k}})$, and $J_{\mathrm{A}}(\Omega)=\sum_{\bm{k}}\delta(\Omega-E_{1\bm{k}}-E_{2\bm{k}})$, where $E_{1\bm{k}}$ and $E_{2\bm{k}}$ are positive eigenenergies of Bogoliubov quasiparticles ($E_{1\bm{k}}\geq E_{2\bm{k}} \geq 0$) in the superconducting state. The subscript ${\rm A}$ and ${\rm B}$ correspond to the type A and type B transitions in Fig.~\ref{fig:transition_image}, respectively. We see that $\sigma^{y;yx}_{\mathrm{PC}}$ and $J_{\mathrm{B2}}(\Omega)$ share the peak position. Thus, it is indicated that the frequency dependence of the photocurrent conductivity is roughly determined by the joint density of states $J_{\mathrm{B2}}(\Omega)$. This result implies that the type B transition ($E \leftrightarrow -E$) dominantly contributes to the photocurrent generation under the circularly-polarized light~\cite{Xu2019}.
The excitation channel $E \leftrightarrow -E$ is unique to the superconducting state and does not have the counterpart in limit of the zero pair potential, namely, the normal state.
The diverging response also appears in the subgap regime due to the anomalous photocurrent mechanism and is almost unrelated to the joint density of states.

\begin{figure}[tbp]
 \includegraphics[width=\linewidth]{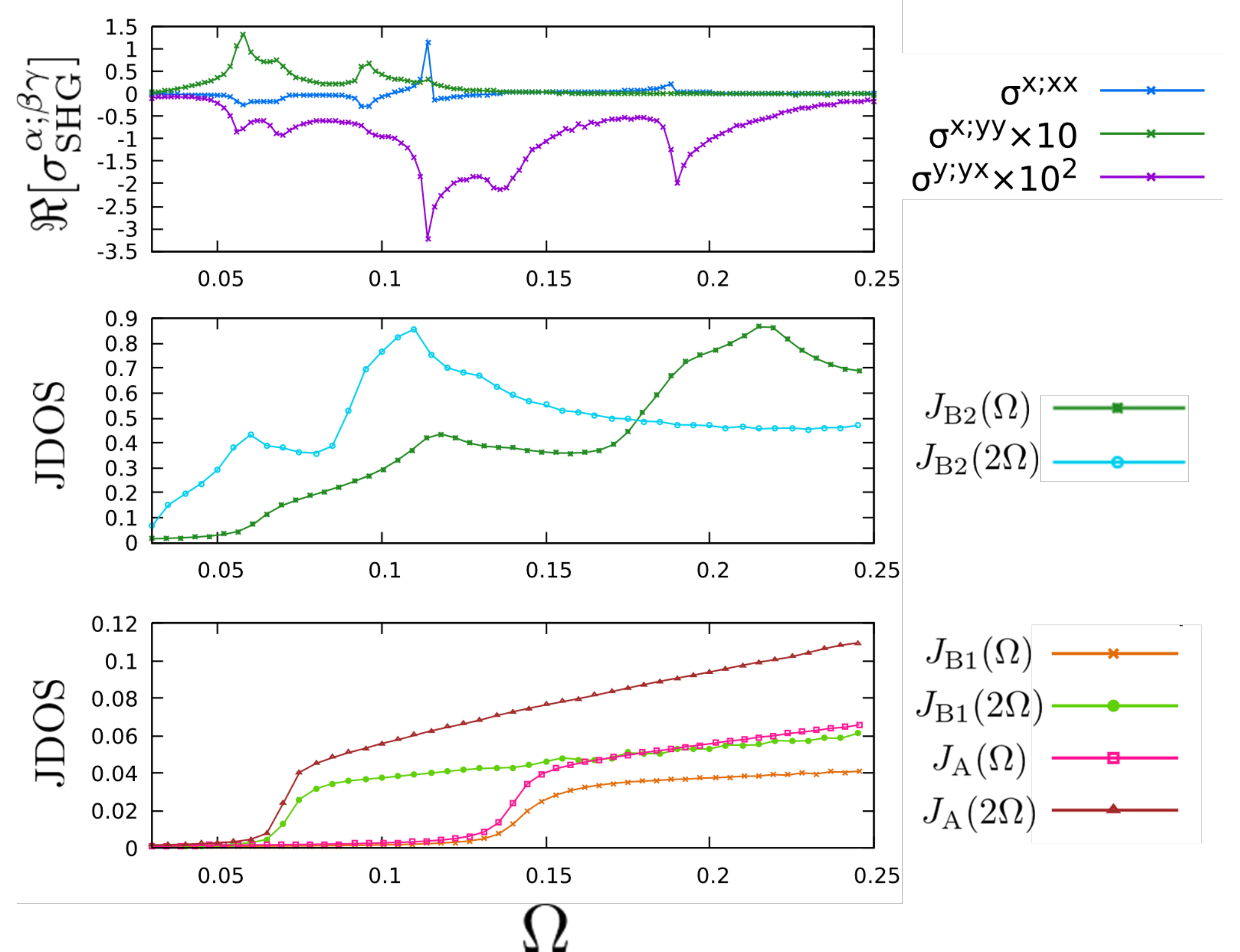}
 \caption{(Upper panel) frequency dependence of the real part of the second harmonic generation. Joint density of states (middle panel)  $J_{\mathrm{B1}}(\Omega)$ and $J_{\mathrm{B1}}(2\Omega)$, and  (lower panel) $J_{\mathrm{B2}}(\Omega)$, $J_{\mathrm{B2}}(2\Omega)$, $J_{\mathrm{A}}(\Omega)$, and $J_{\mathrm{A}}(2\Omega)$. Parameters are the same as the No.1 case in Table~\ref{tab:s_p_model_para}. }
 \label{fig:s_p_model_DOS_SHG_Re}
\end{figure}

In Fig.~\ref{fig:s_p_model_DOS_SHG_Re} we show the frequency dependence of the real part of the second harmonic generation $\sigma^{x;xx}_{\mathrm{SHG}}$, $\sigma^{x;yy}_{\mathrm{SHG}}$, $\sigma^{y;yx}_{\mathrm{SHG}}$ with the joint density of states. Contrary to the photocurrent, not only the transition $\Delta E \sim \Omega$ but also $\Delta E \sim 2\Omega$ is important for the second harmonic generation, where $\Delta E$ is a difference of energies between transition bands~\cite{Xu2019}. Thus, we also show $J_{\mathrm{B1}}(2\Omega)$, $J_{\mathrm{B2}}(2\Omega)$, and $J_{\mathrm{A}}(2\Omega)$, which correspond to the joint density of states about $\Delta E \sim 2\Omega$ transitions. %Joint densities of states
We can find that $J_{\mathrm{B2}}(\Omega)$ and $J_{\mathrm{B2}}(2\Omega)$ share some peak positions with second harmonic generation coefficients. It is indicated that the frequency dependence of the second harmonic generation is roughly determined by not only $J_{\mathrm{B2}}(\Omega)$ but also $J_{\mathrm{B2}}(2\Omega)$. From these comparisons with the joint density of states, we conclude that $E\leftrightarrow -E$ (type B) transition, which is unique to the superconducting state, dominantly contributes to the second-order nonlinear optical responses. 

\subsection{Extended $s$+$p$ model}
\label{subsec:extened_s+p_model}

\begin{figure*}[tbp]
 \includegraphics[width=\linewidth]{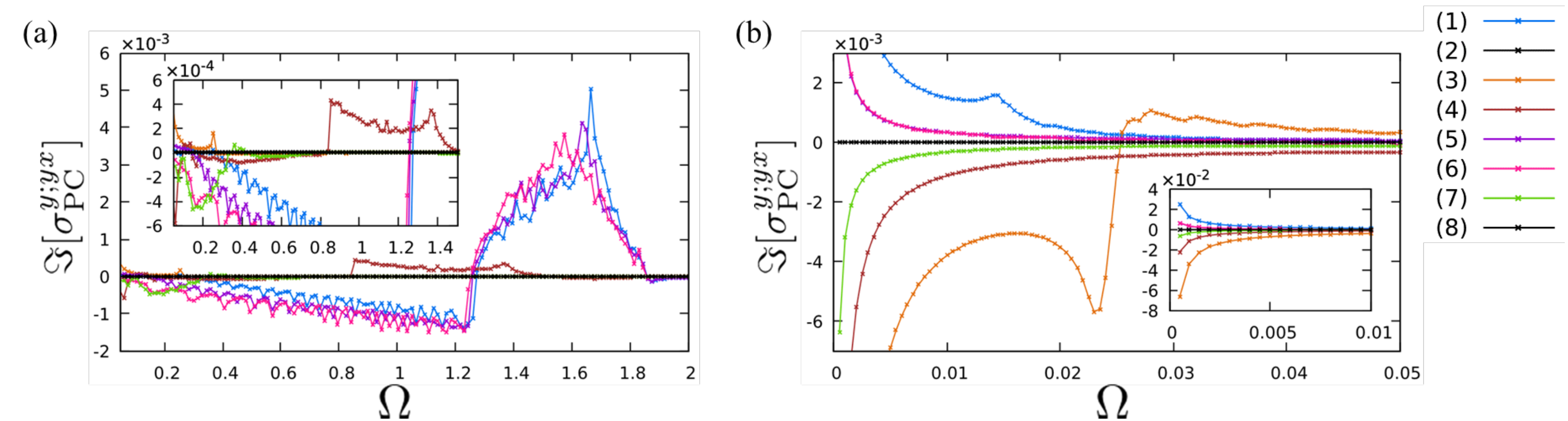}
 \caption{Frequency dependence of the imaginary part of photocurrent conductivity $\sigma^{y;yx}_{\mathrm{PC}}$ in the extended $s$+$p$ model for various parameter sets.
 %with varying the parameter $\alpha_{1}\cdots\alpha_{3}$, $\psi_{0}$, and $d_{1}\cdots d_{3}$. 
 Table~\ref{tab:extended_s_p_model_para} shows the parameter sets of the cases (1)-(8). Panel (a) shows the frequency range $0.05 < \Omega <2.0$, and panel (b) shows $0< \Omega < 0.05$.}
 \label{fig:extend_s_p_model_para}
\end{figure*}

\begin{table*}[tbp]
\centering
\begin{tabular}{|c|c|c|c|c|c|c|c|c|c|c|}
\hline
No. &  $(\alpha_{1},\alpha_{2},\alpha_{3})$ & $\psi_{0}$ & $(d_{1}, d_{2}, d_{3})$ & $\bm{d}_{\bm{k}}\cdot\bm{g}_{\bm{k}}$ & $\bm{d_{k}}\times\bm{g_{k}}$ & $(\partial_{x}\bm{g_{k}}\times\partial_{y}\bm{g_{k}})\cdot\hat{\bm{g}}_{\bm{k}}$ & $(\partial_{x}\bm{g_{\bm{k}}}\times\partial_{y}\bm{g_{\bm{k}}})\cdot\hat{\bm{d}}_{\bm{k}}$& $\Delta_{\mathrm{intra}}$ & $\Delta_{\mathrm{inter}}$ & response \\
 \hline
1 & (0.8, 0.2, 0.6) & 0.08 & (0.01, 0.05, 0.06) & nonzero & nonzero & nonzero & nonzero & nonzero & nonzero & nonzero \\
\hline
2 & (0, 0.6, 0.4) & 0.02 & (0, 0.08, 0.07) & nonzero & nonzero & zero & zero & nonzero & nonzero & zero \\
\hline
3 & (0.4, 0, 0.3) & 0.06 & (0.05, 0.09, 0.07) & nonzero & nonzero & zero & zero & nonzero & nonzero & nonzero \\
\hline
4 & (0.8, 0.5, 0) & 0.06 & (0.01, 0.03, 0.04) & nonzero & nonzero & zero & nonzero & nonzero & nonzero & nonzero\\
\hline
5 & (0.8, 0.2, 0.6) & 0.06 & (0.04, 0.01, 0.03) & nonzero & zero & nonzero & nonzero & nonzero & zero & nonzero \\
\hline
6 & (0.8, 0.2, 0.6) & 0 & (0, 0, 0) & zero & zero & nonzero & undefined & zero & zero & nonzero \\
\hline
7 & (0.2, 0.8, 0) & 0.04 & (0, 0, 0.03) & zero & nonzero & zero & nonzero & nonzero & nonzero & nonzero \\
\hline
8 & (0.2, 0.8, 0) & 0 & (0, 0, 0.03) & zero & nonzero & zero & nonzero &  zero & nonzero & zero \\
\hline
\end{tabular}
\caption{
The parameter set $(\alpha_{1}, \alpha_2, \alpha_{3})$, $\psi_{0}$, and $(d_{1}, d_2, d_{3})$ adopted in Fig.~\ref{fig:extend_s_p_model_para} and existence or non-existence of nonlinear optical responses (first right column). %When $\bm{d}_{\bm{k}}\cdot\bm{g}_{\bm{k}}$ or $\psi_0$ is nonzero, 
%the pair potential has an intraband component. %The second right column shows whether the SC Berry curvature factor is zero or nonzero.
$\bm{d}_{\bm{k}}\cdot\bm{g}_{\bm{k}}$ and $\bm{d_{k}}\times\bm{g_{k}}$ represent the intraband and interband components in the spin-triplet pair potential $\bm{d_{k}}$. The fifth and fourth right columns show whether the normal and SC Berry curvature factors are zero or nonzero, respectively.}
\label{tab:extended_s_p_model_para}
\end{table*}

Next, the photocurrent conductivity and second harmonic generation in the extended $s$+$p$ model are presented. An essential difference between the $s$+$p$ and extended $s$+$p$ models is the normal Berry curvature factor, which can be nonzero in the extended $s$+$p$ model. Therefore, the nonlinear optical responses may appear in the normal state of extended $s$+$p$ model (No.6 case in Fig.~\ref{fig:extend_s_p_model_para} and Table~\ref{tab:extended_s_p_model_para}), in contrast to the $s$+$p$ model.
The normal and SC Berry curvature factors in the extended $s$+$p$ model are given by
\begin{align}
&\partial_{x}\bm{g_{k}}\times\partial_{y}\bm{g_{k}}\cdot\hat{\bm{g_{k}}} \notag \\
&\quad \propto \alpha_{1}\alpha_{2}\alpha_{3}\cos k_{x}(\sin 2k_{y} \cos k_{y} - 2\sin k_{y} \cos 2k_{y}), \\
&\partial_{x}\bm{g_{k}}\times\partial_{y}\bm{g_{k}}\cdot\hat{\bm{d_{k}}}\notag \\ 
& \quad\propto \alpha_{2}\cos k_{x}\times(d_{1}\alpha_{3}\sin 2k_{y}\cos k_{y} - 2d_{3} \alpha_{1}\sin k_{y} \cos 2k_{y}).
\end{align}
When none of $\alpha_{1}$, $\alpha_2$, $\alpha_{3}$ are zero, the normal Berry curvature factor is finite.

Figure \ref{fig:extend_s_p_model_para} plots the photocurrent conductivity $\sigma^{y;yx}_{\mathrm{PC}}$ with varying the parameters $(\alpha_{1}, \alpha_2, \alpha_{3})$, $\psi_{0}$, and $(d_{1}, d_2, d_{3})$. The parameter sets and whether $\sigma^{y;yx}_{\mathrm{PC}}$ is zero or nonzero are summarized in Table~\ref{tab:extended_s_p_model_para}. First, the photocurrent conductivity can be finite even when the SC Berry curvature factor vanishes (No.3 case in Table~\ref{tab:extended_s_p_model_para}), although the SC Berry curvature factor is essential in the $s$+$p$ model. %for the second-order nonlinear responses. In the No.3 case whose Normal and SC Berry curvature factor is zero, however, the second-order nonlinear responses are nonzero
A finite SC Berry curvature factor is not necessarily required for the nonlinear optical responses in the extended $s$+$p$ model.
Thus, implication from the analysis of the injection current (Sec.~\ref{subsec:condition_Einj}) does not necessarily apply.
This is probably because the extended $s$+$p$ model is more complicated than the $s$+$p$ model. 

\begin{figure}[tbp]
 \includegraphics[width=\linewidth]{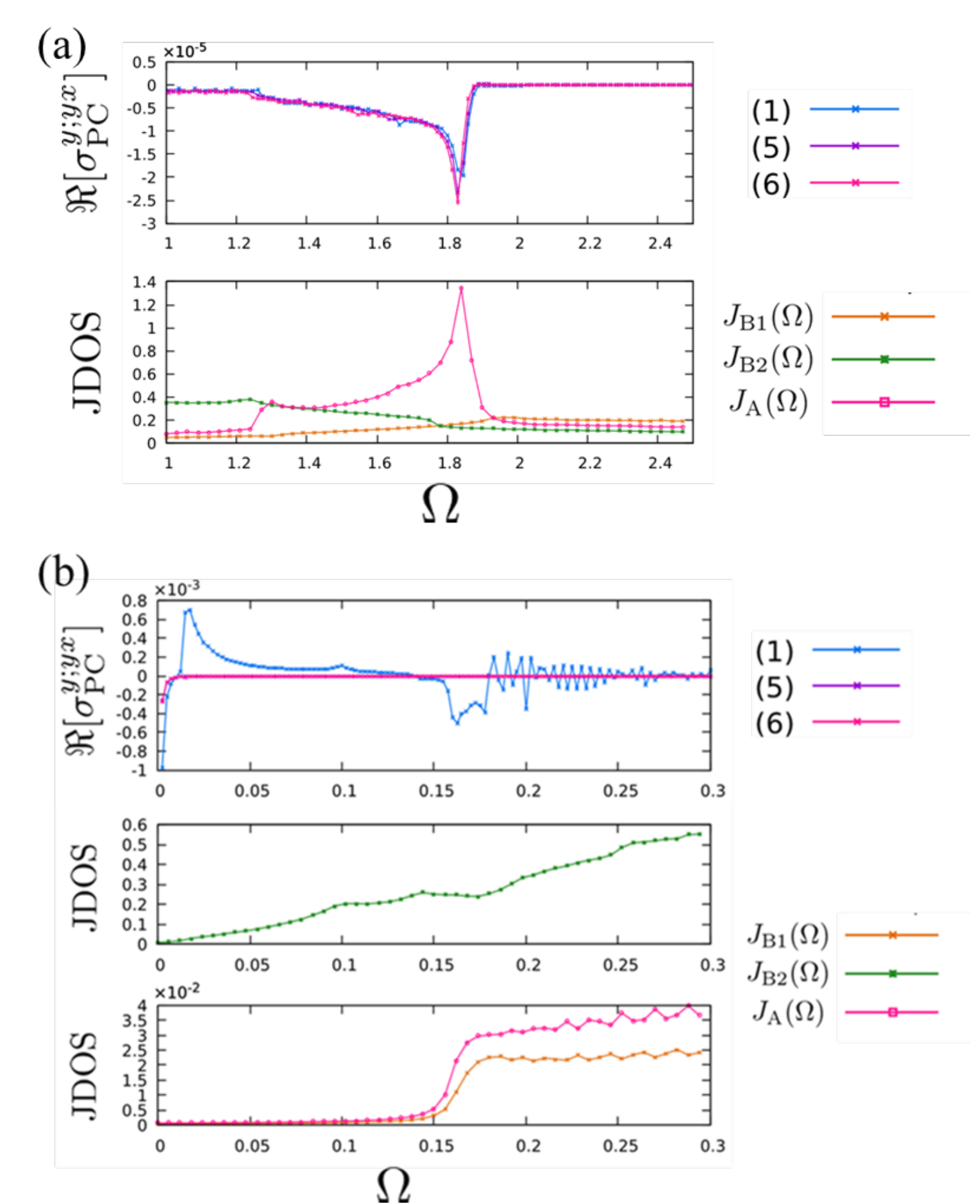}
 \caption{(a) The upper panel shows the real part of the photocurrent conductivity $\sigma^{y;yx}_{\mathrm{PC}}$, and the lower panel shows the joint density of states for each transition process in the extended $s$+$p$ model. A high frequency region $1< \Omega < 2.5$ is shown. (b) The same quantities are plotted in a low frequency region $0< \Omega < 0.3$.  %frequency dependence of (upper panel) the real part of the photocurrent conductivity $\sigma^{y;yx}_{\mathrm{PC}}$ and joint density of states (middle panel) $J_{\mathrm{intra2}}(\Omega)$, (lower panel) $J_{\mathrm{intra1}}(\Omega)$, and $J_{\mathrm{inter}}(\Omega)$. 
 The parameter sets in the No.1, 5, and 6 cases of Table~\ref{tab:extended_s_p_model_para} are adopted for the photocurrent conductivity. The normal Berry curvature factor is finite in these parameter sets.
 The parameters for the joint density of states are the No.1 case. %\ref{tab:extended_s_p_model_para}.
 }
 \label{fig:DOS_Re_yyx_extend_s_p_comp}
\end{figure}

Next, we focus on the cases with finite normal Berry curvature factor and compare the photocurrent conductivity of the No.1, 5, and 6 cases. The parameters for the spin-orbit coupling $(\alpha_{1}, \alpha_2, \alpha_{3})$ are equivalent in these cases. The No.6 case represents the normal state because the pair potential is zero, while the No.1 and No.5 cases represent the superconducting state. The pair potential in the No.1 case consists of intraband and interband components, while interband pairing is absent in the No.5 case. For additional information, in Fig.~\ref{fig:DOS_Re_yyx_extend_s_p_comp}, we show the real part of photocurrent conductivity $\sigma^{y;yx}_{\mathrm{PC}}$ in these cases and the joint density of states in the No.1 case. %The pair potential in No.6 is zero and photocurrent of normal state appears due to nonzero normal Berry curvature factor. 
Figures~\ref{fig:DOS_Re_yyx_extend_s_p_comp}~(a) and \ref{fig:extend_s_p_model_para} reveal that the photocurrent conductivity is almost the same in the frequency region $1\leq \Omega\leq 2.5$. Furthermore, we see that the real part of $\sigma^{y;yx}_{\mathrm{PC}}$ and $J_{\mathrm{A}}(\Omega)$ share the peak position. Thus, it is indicated that the frequency dependence of the photocurrent conductivity is roughly determined by the joint density of states $J_{\mathrm{A}}(\Omega)$, in contrast to the $s$+$p$ model discussed in the previous subsection. These results imply that the photocurrent generation %of No.1 and 5 
for $1\leq \Omega\leq 2.5$ 
mainly stems from the normal photocurrent mechanism even in the superconducting state, which arises from $E_{i}\leftrightarrow -E_{j} \,\, (i\neq j)$ transition (type A in Fig. \ref{fig:transition_image}). %dominantly in the superconducting state.

On the other hand, we also find that nonlinear optical responses are drastically modified by superconductivity. Figure~\ref{fig:DOS_Re_yyx_extend_s_p_comp}~(b) shows a sizable photocurrent conductivity %$\sigma^{y;yx}_{\mathrm{PC}}$ 
around $\Omega = 0.015$ only in the No.1 case. Such a large photocurrent generation does not appear in the No.6 case, and therefore, it is unique to the superconducting state. %The photocurrent conductivity of the 
Comparison with the No.5 case, in which the pair potential does not have an interband component, %is almost zero in the low-frequency region. It is 
implies that interband pairing is essential for the photocurrent generation unique to the superconducting state. %\YYS{It is difficult to identify which transition contributes to the photocurrent generation in the low-frequency region because we cannot see the peak shared by photocurrent conductivity and joint density of states.}

\begin{figure}[tbp]
 \includegraphics[width=\linewidth]{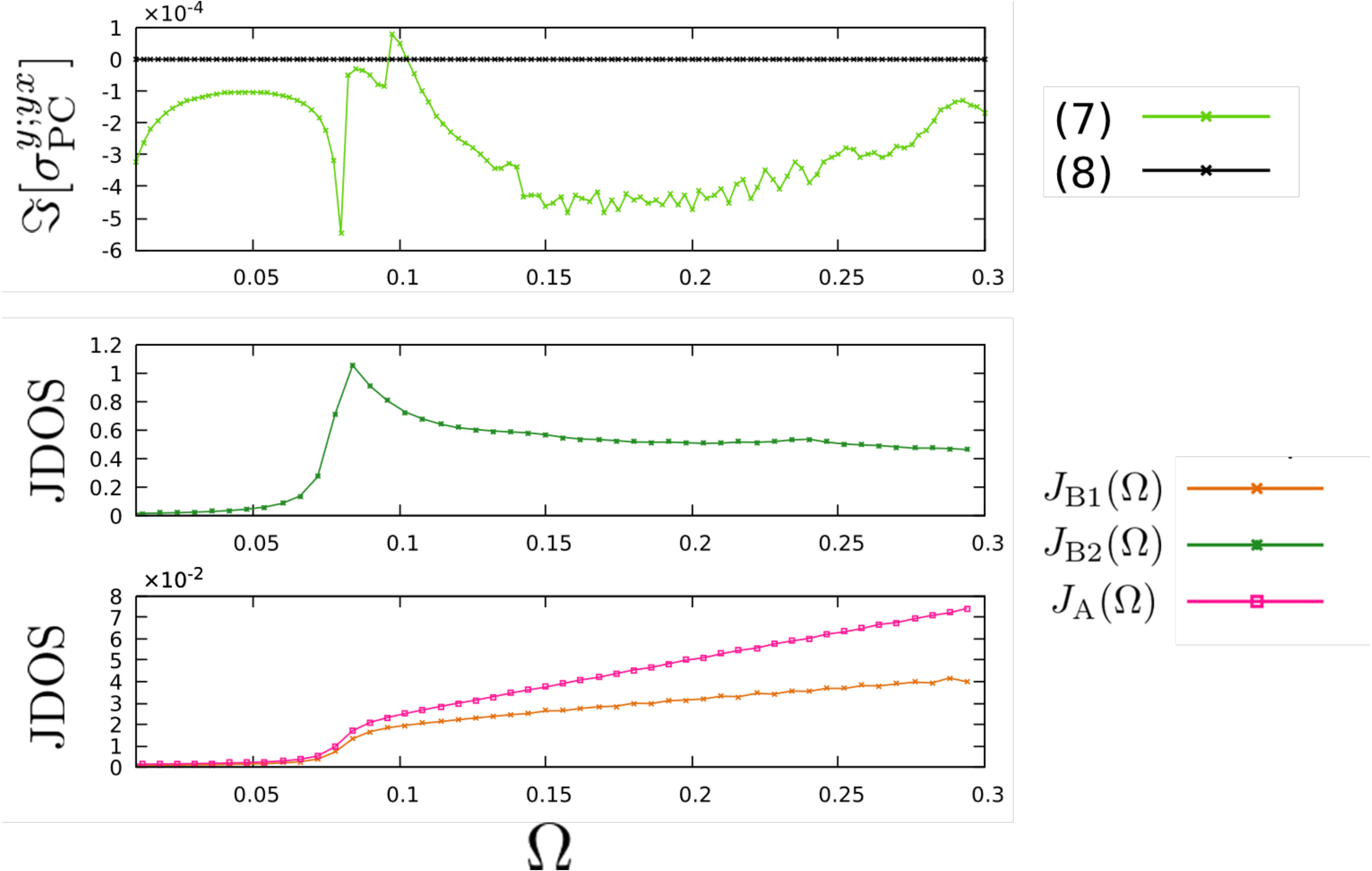}
 \caption{(Upper panel) the imaginary part of the photocurrent conductivity $\sigma^{y;yx}_{\mathrm{PC}}$. (Middle panel) joint density of states $J_{\mathrm{B2}}(\Omega)$. (Lower panel) $J_{\mathrm{B1}}(\Omega)$ and $J_{\mathrm{A}}(\Omega)$. The parameters in the No.7 case are adopted for the joint density of states. %The parameter sets in the No.1, 6, and 7 cases are adopted for the photocurrent conductivity.
% The parameters of joint density operator shown in this figure are same as ones of No.7 case. 
 Photocurrent conductivity is calculated for the No.7 and 8 cases %The parameters of (7) and (8) are summarized 
 in Table~\ref{tab:extended_s_p_model_para}, where the normal Berry curvature factor is zero.}
 \label{fig:inter_DOS_extend_s_p_comp}
\end{figure}

Finally, we compare %the photocurrent conductivity of 
the No.7 and 8 cases, in which the normal Berry curvature factor is zero. 
In Figs.~\ref{fig:extend_s_p_model_para} and \ref{fig:inter_DOS_extend_s_p_comp}, we see that the photocurrent conductivity $\sigma^{y;yx}_{\mathrm{PC}}$ vanishes in the No.8 case, while finite in the No.7 case.
The difference between the two cases is the presence or absence of the spin-singlet pair potential $\psi_0$.
In both cases, the spin-triplet pair potential $\bm{d}_{\bm k}$ does not contain an intraband component. 
Therefore, disappearance of the photocurrent conductivity in the No.8 case is attributed to the absence of the intraband pairing. 
%and  %The difference in these cases is only whether the spin-singlet is zero or nonzero. 
It is again implied that the coexistence of intraband pairing and interband pairing is essential for the second-order nonlinear optical responses in superconductors when the responses are absent in the normal state.
%\YYS{when the normal Berry curvature factor is zero.} 
In Fig.~\ref{fig:inter_DOS_extend_s_p_comp}, we compare the imaginary part of $\sigma^{y;yx}_{\mathrm{PC}}$ with $J_{\mathrm{B2}}(\Omega)$ and see that they share the peak position. %Thus, it is indicated that the frequency dependence of the photocurrent conductivity is roughly determined by the joint density of state $J_{\mathrm{intra2}}(\Omega)$. 
Thus, the photocurrent generation in the No.7 case is attributable to the type B transition unique to the superconducting state. 

\section{Discussion}
\label{sec:discussion}

In this paper, we have investigated the conditions where the second-order nonlinear optical responses appear in the superconducting state. This section summarizes the results of photocurrent and second harmonic generations in the models for noncentrosymmetric superconductors, namely, the $s$+$p$ and extended $s$+$p$ models. 

First, we discuss the relation between the nonlinear responses and the Berry curvature factors. In the superconducting state of the $s$+$p$ model, a nonzero SC Berry curvature factor is essential for the second-order nonlinear responses to appear. Comparison with the joint density of states shows that the photocurrent and second harmonic generations stem from the type B transition. On the other hand, finite Berry curvature factors are not necessarily required for the nonlinear responses in the extended $s$+$p$ model. In the No.3 case of Table~\ref{tab:extended_s_p_model_para}, where the normal and SC Berry curvature factors vanish, the photocurrent and second harmonic generation occur. However, the nonlinear responses %in the extended $s$+$p$ model 
are not independent of the Berry curvature factors. In the No.2 case, indeed, the second-order nonlinear responses disappear probably because the Berry curvature factors are zero. Therefore, we conclude that the second-order nonlinear optical responses unique to the superconducting state are closely related to the SC Berry curvature factor.

Next, we discuss the effects of intraband and interband pairing on the second-order nonlinear responses. In the $s$+$p$ model, the second-order nonlinear responses of normal states are absent. Moreover, the photocurrent and second harmonic generations are absent in the superconducting states when the pair potential does not contain either the intraband or interband components. The same results are obtained in the extended $s$+$p$ model %both the intraband and interband pairing is required for the second-order nonlinear responses 
when the normal Berry curvature factor is zero. When the normal Berry curvature factor is finite, then the second-order nonlinear responses appear in the superconducting states without interband pairing. However, the responses are almost equivalent to those in the normal state. Therefore, both intraband and interband pairing are necessary for the emergence of the second-order nonlinear optical responses unique to the superconducting state. 

We suppose that the type B transition mainly contributes to the optical response unique to the superconducting state. In this paper, we have shown that the joint density of states for the type B transition and the second-order nonlinear conductivities share the peaks for the parameter set No.1 of the $s$+$p$ model and the No.7 of the extended $s$+$p$ model. This finding implies that the type B transition contributes to the superconducting nonlinear response. Considering the condition where the superconducting nonlinear response appears, we think that the coexistence of intraband and interband pairing is important for the type B transition in our models. In addition, it is known that a unique contribution to the optical linear response through the type B transition is forbidden in the $\mathcal{T}$-symmetric superconductor when the pair potential contains only the intraband components~\cite{Ahn2021}. Further analytical discussion about the correspondence of nonlinear responses and the microscopic transition process is left for future work.

From the results obtained above, it is expected that a mixture of intraband and interband pairing is important to maximize the second-order nonlinear optical response. To verify this expectation, we calculate the photocurrent conductivity in the $s$+$p$ model with varying the ratio of intraband and interband pairing. We define $\Delta_{\mathrm{intra}}$ and $\Delta_{\mathrm{inter}}$ to represent the intraband and interband pairing components in the pair potential,
%respectively. We represent the components of intraband and interband pairing by 
\begin{equation}
\Delta_{\mathrm{intra}}=[\psi_{\bm{k}}+\bm{d}_{\parallel \bm{k}}\cdot\bm{\sigma}]i\sigma_{y}, \quad \Delta_{\mathrm{inter}}=[\bm{d}_{\perp \bm{k}}\cdot\bm{\sigma}]i\sigma_{y},
\end{equation}
and take the parameters as
\begin{gather}
\bm{d}_{\parallel \bm{k}}=(0.03\sin k_{y},0.02\sin k_{x},0.04\sin k_{y}), \\
\bm{d}_{\perp \bm{k}}=(0.08\sin k_{y}, 0, -0.06\sin k_{y}), \\
\psi_{\bm{k}}=0.06, \quad (\alpha_{1}, \alpha_{2}, \alpha_{3})=(0.3, 0.2, 0.4).
\end{gather}
The other parameters are the same as those in Sec.~\ref{sec:result}. %Secs.~\ref{subsec:s+p_model} and \ref{subsec:extened_s+p_model}.
\begin{figure}[tbp]
 \includegraphics[width=\linewidth]{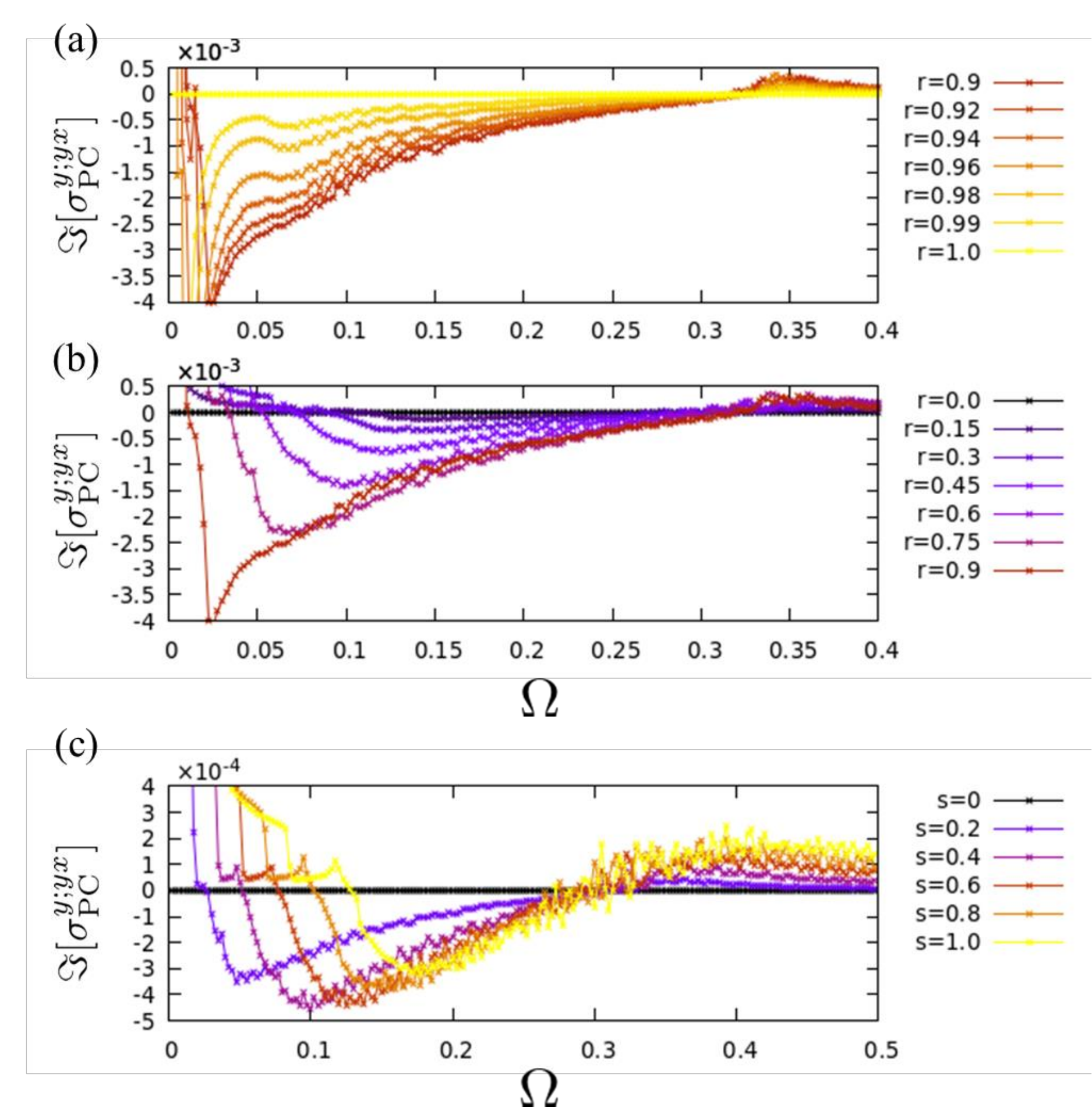}
 \caption{Change in the photocurrent conductivity $\Im [\sigma^{y;yx}_{\mathrm{PC}}]$ in the $s$+$p$ model with varying the parameter $r$ of the ratio of intraband and interband pairing. (a) $0.9\leq r \leq 1.0$ and (b) $0\leq r \leq 0.9$. (c) %The change in the frequency dependence of the imaginary part of photocurrent conductivity 
 $\Im [ \sigma^{y;yx}_{\mathrm{PC}}]$ in the $s$+$p$ model with various amplitudes of the pair potential. The pair potentials adopted in the figure are described in the main text.}
 \label{fig:intra_inter_trans}
\end{figure}
Figures~\ref{fig:intra_inter_trans}~(a) and \ref{fig:intra_inter_trans}~(b) plot the imaginary part of photocurrent conductivity $\sigma^{y;yx}_{\mathrm{PC}}$ with varying the parameter $r$ for the ratio of intraband and interband pairing, which defines the pair potential
\begin{equation}
\Delta_{\bm k} = (1-r)\Delta_{\mathrm{intra}}+r\Delta_{\mathrm{inter}}.
\end{equation}
At $r=0$ and $r=1$, the photocurrent generation vanishes as expected because the mixture of intraband and interband pairing is absent. The photocurrent conductivity is maximized near $r\simeq 0.9$ and drastically changes around $r=0.9$.
%shows a sharp peak as a function of $r$.}

We also show Fig.~\ref{fig:intra_inter_trans}~(c) to clarify the dependence on the magnitude of the pair potential. 
%plots the frequency dependence of the imaginary part of photocurrent conductivity $\sigma^{y;yx}_{\mathrm{PC}}$ with varying the amplitude parameter $s$ of pair potential, which is defined as
Here, we adopt $\Delta_{\bm k} = s\tilde{\Delta}_{\bm k}$ with %We %take the parameters of 
$\tilde{\Delta}_{\bm k}=[\tilde{\psi}_{\bm{k}}+\tilde{\bm{d}}_{\bm{k}}\cdot\bm{\sigma}]i\sigma_{y}$ and
\begin{equation}
\tilde{\psi}_{\bm{k}}=0.06, \quad \tilde{\bm{d}}_{\bm{k}}=(0.07\sin k_{y},0.02\sin k_{x},0.01\sin k_{y}).
\end{equation}
%Other parameters are the same as ones of Figure \ref{fig:intra_inter_trans} (a) and (b). 
Note that the normal Berry curvature factor vanishes in this case.
The photocurrent conductivity is zero at $s=0$ because the pair potential is zero, and the model represents the normal state. With increasing the amplitude of the pair potential, the behavior of the photocurrent conductivity shifts in parallel to the higher frequency side. 
This is reasonable because the magnitude of the superconducting gap increases with the pair potential. 
%\YYS{The results of Figure \ref{fig:intra_inter_trans} imply that the changes in the frequency dependence of photocurrent conductivity are different between varying the proportion of intraband and interband pairing and varying the amplitude of pair potential. Clarification of the difference is left for future.}

\section{Conclusion}
\label{sec:conclusion}
This paper elaborated on the second-order nonlinear optical responses of noncentrosymmetric superconductors. In particular, we performed detailed analysis of typical models for superconductors with $\mathcal{T}$ symmetry and clarified microscopic conditions for the nonlinear responses beyond the symmetry constraints. Our analytic calculations imply that the normal and SC Berry curvature factors and the mixture of intraband and interband pairing are closely related to the presence of photocurrent and second harmonic generations in superconductors. %The photocurrent conductivity and second harmonic generation coefficient are numerically calculated in the $s$+$p$ model and extended $s$+$p$ model.
This implication has been justified in the numerical calculations. 

In the numerical analysis of the $s$+$p$ model and extended $s$+$p$ model, first, we revealed that the second-order nonlinear responses are suppressed when the Berry curvature factors are zero. Especially, the SC Berry curvature factor is related to the contributions characteristic of the superconducting state. Second, we clarified that the second-order nonlinear responses unique to the superconducting state originate from a mixture of intraband and interband pairing. %Parity-breaking of pair potential is not necessarily required because the spin-triplet $\bm{d}$ can consist of both the intraband and interband pairing. 
Finally, %we investigated the changes in the frequency dependence of photocurrent conductivity with varying proportions of the intraband and interband pairing. We can
we found that the photocurrent generation can be maximized by tuning the mixture of intraband and interband pairing. Combining the analytical and numerical results, we conclude that the SC Berry curvature factor and the mixture of intraband and interband pairing are essential for the second-order nonlinear optical responses in $\mathcal{T}$-symmetric superconductors.

%In $\mathcal{T}$ symmetric superconductors, the second-order nonlinear responses are strongly affected by the proportion of the intraband and interband pairing. Thus, the changes in the frequency dependence of nonlinear responses will help investigate how the pair potential changes. 
Among various mechanisms to break space inversion symmetry in superconductors, the most typical and ubiquitous class is the noncentrosymmetric superconductors lacking the inversion symmetry in the crystal structure. Examples range from heterostructures~\cite{Reyren2007,Ueno2008,Saito2016,Lu2015,Xi2016,delaBarrera2018} to bulk compounds~\cite{Bauer2012,Smidman2017}. Unless an external field such as the magnetic field is applied, this class of superconductors preserves $\mathcal{T}$ symmetry. Thus, our study analyzing such superconductors would be helpful for future developments in superconductivity and nonlinear optics. For example, the nonlinear optical responses may signify the parity mixing of Cooper pairs. Although the even-parity spin-singlet pair potential and odd-parity spin-triplet one can coexist in noncentrosymmetric superconductors~\cite{Gor'kov2001,Frigeri2004,Bauer2012,Smidman2017}, an unambiguous observation has been awaited for a long time. This is because most of the characteristic properties of noncentrosymmetric superconductors, such as boosted upper critical field~\cite{Kimura2007,Settai2008,Saito2016,Lu2015,Xi2016,delaBarrera2018}, arise from the spin-momentum locking even when the parity mixing does not occur. On the other hand, as we discussed in this paper, the superconducting nonlinear optical responses are sensitive to the structure of pair potential and need spin-triplet pairing. Because most superconductors are spin-singlet superconductors, the observation of nonlinear optical responses would indicate the presence of spin-triplet Cooper pairs originating from the parity mixing phenomenon.

From the engineering perspective, the ratio of intraband pairing to interband pairing is an important parameter to control the nonlinear responses unique to the superconducting state. 
The intraband pairing is expected to be dominant since the interband pairing hardly contributes to the thermodynamic stability of superconductivity. However, the interband pairing inevitably appears when we adopt a strong coupling theory for superconductivity~\cite{Bauer2012,Yanase2008}, and therefore, it is expected to be finite in realistic superconductors. In particular, it has been proposed that a significant interband pairing appears when the spin-orbit coupling has topological defects in the momentum space, such as Weyl points~\cite{Bauer2012,Yanase2008}. Based on these arguments, superconducting WTe$_2$ is predicted to be a candidate for significant nonlinear optical responses. The symmetry of the models studied in this paper is consistent with the bilayer WTe$_2$~\cite{Ma2019}. The study of this class of materials~\cite{Fetami2018} will be interesting for future research. 
Furthermore, it is expected that the nonlinear responses, which are not uncovered in this paper, may appear in $\mathcal{T}$ symmetry breaking superconductors~\cite{Watanabe2022}. 
Thus, exploring other superconducting nonlinear responses will also be a promising route for future research.

%%%%%%%%%% Acknowledgments %%%%%%%%%%
\begin{acknowledgments}
The authors are grateful to A. Daido for fruitful discussions.
This work was supported by JSPS KAKENHI (Grants Nos. JP18H05227, JP18H01178, JP20H05159, JP21K18145, JP21J00453, JP22H01181) and SPIRITS 2020 of Kyoto University.

\end{acknowledgments}

\appendix
\onecolumngrid

\section{Photocurrent conductivity in the normal state}

In this section, we derive the photocurrent conductivity of the two-dimensional $\mathcal{T}$ symmetric system in the normal state. %when the normal Berry curvature factor is zero. 
The Hamiltonian is given by
\begin{equation}
H_\text{N}(\bm{k})=\xi_{\bm{k}}+\bm{g_{k}}\cdot\bm{\sigma}.
\end{equation}
The first term is kinetic energy measured from a chemical potential $\mu$ and the second term is an antisymmetric spin-orbit coupling. As follows, we omit the subscript $\bm{k}$ for simplicity. The normal photocurrent conductivity consists of the four contributions
\begin{equation}
\sigma_{\mathrm{N}}=\sigma_{\mathrm{BCD}}+\sigma_{\mathrm{IFS(E)}}+\sigma_{\mathrm{Einj}}+\sigma_{\mathrm{shift}},
\end{equation}
which are termed Berry curvature dipole term, intrinsic Fermi surface term, electric injection current term, and shift current term, respectively~\cite{Watanabe2021, Watanabe2022}. Each term is formulated as
\begin{align}
\label{eq:app_ISF}
\sigma^{\alpha;\beta\gamma}_{\mathrm{ISF(E)}}  &= \frac{i}{4}\sum_{a\neq b}\Omega^{\beta\gamma}_{ab}\mathrm{P}\frac{1}{\Omega-E_{ab}}\partial_{\alpha}f_{ab},  \\
\label{eq:app_BCD}
\sigma^{\alpha;\beta\gamma}_{\mathrm{BCD}}  &= -\frac{i}{2\Omega}\sum_{a}(\epsilon_{\alpha\beta\delta}\partial_{\gamma}\Omega^{\delta}_{a}-\epsilon_{\alpha\gamma\delta}\partial_{\beta}\Omega^{\delta}_{a})f_{a},\\
\label{eq:app_Einj}
\sigma^{\alpha;\beta\gamma}_{\mathrm{Einj}} &= -\frac{i\pi}{4\eta}\sum_{a\neq b}(J^{\alpha}_{aa}-J^{\alpha}_{bb})\Omega^{\beta\gamma}_{ba}f_{ab}\delta(\Omega-E_{ba}), \\
\label{eq:app_shift}
\sigma^{\alpha;\beta\gamma}_{\mathrm{shift}} &= -\frac{\pi}{4}\sum_{a\neq b}\Im\left[\left[D_{\alpha}\xi^{\beta}\right]_{ab}\xi^{\gamma}_{ba} + \left[D_{\alpha}\xi^{\gamma}\right]_{ab}\xi^{\beta}_{ba}\right]f_{ab}\delta(\Omega-E_{ba}),
\end{align}
where we defined geometric quantities such as the Berry connection $\xi^{\alpha}_{ab}$, Berry curvature $\Omega^{\alpha\beta}_{ab}$, and the covariant derivative $D_{\alpha}$.
The sign `P' in Eq.~\eqref{eq:app_ISF} means the principal integral over the frequency $\Omega$.
We also introduced the Berry curvature for the $a$-th band as 
\begin{equation}
\Omega^{\gamma}_{a}=\frac{1}{2}\sum_{b(\neq a)}\epsilon_{\alpha\beta\gamma}\Omega^{\alpha\beta}_{ab}.
\end{equation}
Here, we derive the Berry curvature $\Omega^{\alpha\beta}_{ab}$. The Hamiltonian in the coordinate $\hat{z}\parallel \bm{g}$ has the form 
\begin{equation}
H(\bm{k})=\mathrm{diag}(E_{+},E_{-})=\mathrm{diag}(\xi+g,\xi-g).
\end{equation}
In %$\hat{z}\parallel \hat{g}$ 
this coordinate, the paramagnetic and diamagnetic current operators are given by
\begin{align}
\label{eq:para}
J^{\alpha}(\bm{k}) &= \partial_{\alpha}H(\bm{k})=
\begin{pmatrix}
\partial_{\alpha}\xi+\partial_{\alpha}g_{z}  &  \partial_{\alpha}g_{x}-i\partial_{\alpha}g_{y}  \\
\partial_{\alpha}g_{x}+i\partial_{\alpha}g_{y}  &  \partial_{\alpha}\xi-\partial_{\alpha}g_{z}
\end{pmatrix},
\\
\label{eq:diam}
J^{\alpha\beta}(\bm{k}) &=\partial_{\alpha\beta}H(\bm{k})=
\begin{pmatrix}
\partial_{\alpha\beta}\xi+\partial_{\alpha\beta}g_{z}  &  \partial_{\alpha\beta}g_{x}-i\partial_{\alpha\beta}g_{y}  \\
\partial_{\alpha\beta}g_{x}+i\partial_{\alpha\beta}g_{y}  &  \partial_{\alpha\beta}\xi-\partial_{\alpha\beta}g_{z}
\end{pmatrix},
\end{align}
where $\partial_{\alpha\beta\cdots}$ means the $\bm{k}$ derivative $\partial_{k_{\alpha}}\partial_{k_{\beta}}\cdots$ for shorthand notation.
% and we use the abbreviation $[\partial_{\alpha}\bm{g}]_{\mu} \rightarrow \partial_{\alpha}g_{\mu}$ for simplicity.
With the Hellmann-Feynman relation, the Berry connection and Berry curvature for $a\neq b$ are given by
\begin{align}
\label{eq:connection}
\xi^{\alpha}_{ab} &=\frac{iJ^{\alpha}_{ab}}{E_{ab}}, \\
\Omega^{\alpha\beta}_{ab} &= -\frac{2}{E_{ab}^{2}}\Im[J^{\alpha}_{ab}J^{\beta}_{ba}].
\end{align}
From Eq.~\eqref{eq:para}, we obtain 
\begin{equation}
\label{eq:Im_J_J}
\Im[J^{\alpha}_{ab}J^{\beta}_{ba}]=\left[(\partial_{\alpha}\bm{g}\times\partial_{\beta}\bm{g})\cdot\bm{e}_{z}\right]
\begin{pmatrix}
0  &  1  \\
-1  &  0
\end{pmatrix}_{ab}.
\end{equation}
Therefore, the Berry curvature is obtained as
\begin{equation}
\label{eq:normal_Berry_curvature}
\Omega^{{\alpha}{\beta}}_{ab} = -\frac{2}{E^{2}_{ab}}\Im[J^{\alpha}_{ab}J^{\beta}_{ba}]=\frac{(\partial_{\alpha}\bm{g}\times\partial_{\beta}\bm{g})_{z}}{2g^{2}}\begin{pmatrix}
0  &  -1  \\
1  &  0
\end{pmatrix}_{ab}
=\frac{(\partial_{\alpha}\bm{g}\times\partial_{\beta}\bm{g})\cdot\hat{\bm{g}}}{2g^{2}}\begin{pmatrix}
0  &  -1  \\
1  &  0
\end{pmatrix}_{ab}.
\end{equation}
We defined the unit vector $\hat{\bm{g}} = \bm{g} / g$.  
When the normal Berry curvature factor is $(\partial_{\alpha}\bm{g}\times\partial_{\beta}\bm{g})\cdot\hat{\bm{g}}=0$, the Berry curvature $\Omega^{{\alpha}{\beta}}_{ab}$ disappears in the normal state. In the two-dimensional systems, only the component $(\partial_{x}\bm{g}\times\partial_{y}\bm{g})\cdot\hat{\bm{g}}=-(\partial_{y}\bm{g}\times\partial_{x}\bm{g})\cdot\hat{\bm{g}}$ can be finite. 

Next, we derive the photocurrent conductivity in the condition $(\partial_{x}\bm{g}\times\partial_{y}\bm{g})\cdot\hat{\bm{g}}=0$. From Eqs.~\eqref{eq:app_ISF}, \eqref{eq:app_BCD}, and \eqref{eq:app_Einj}, we obtain $\sigma_{\mathrm{IFS(E)}}=\sigma_{\mathrm{BCD}}=\sigma_{\mathrm{Einj}}=0$.
Therefore, we have only to evaluate the shift current term. In the two-band system, we obtain
\begin{equation}
\left[D_{\alpha}\xi^{\beta}\right]_{ab}=iJ^{\beta}_{ab}\frac{\Delta^{\alpha}_{ab}}{E_{ab}^{2}}+iJ^{\alpha}_{ab}\frac{\Delta^{\beta}_{ab}}{E_{ab}^{2}}-i\frac{J^{\alpha\beta}_{ab}}{E_{ab}},
\end{equation}
where we define the difference of group velocity $\Delta^{\alpha}_{ab}=J^{\alpha}_{aa}-J^{\alpha}_{bb}$ and obtain as
\begin{equation}
\label{eq:Delta}
\Delta^{\alpha}_{ab}=2\partial_{\alpha}g_{z}
\begin{pmatrix}
0  &  1  \\
-1  &  0  
\end{pmatrix}_{ab}.
\end{equation}
From Eq.~\eqref{eq:connection}, we obtain
\begin{align}
\Im\left(\left[D_{\alpha}\xi^{\beta}\right]_{ab}\xi^{\gamma}_{ba}+\left[D_{\alpha}\xi^{\gamma}\right]_{ab}\xi^{\beta}_{ba}\right)=&\frac{1}{E_{ab}^{3}}\left[\Delta^{\alpha}_{ab}\Im\left(J^{\beta}_{ab}J^{\gamma}_{ba}+J_{ba}^{\beta}J^{\gamma}_{ab}\right)+\Delta^{\beta}_{ab}\Im\left(J^{\gamma}_{ba}J^{\alpha}_{ab}\right)+\Delta^{\gamma}_{ab}\Im\left(J^{\beta}_{ba}J^{\alpha}_{ab}\right) \right]  \notag \\
&-\frac{1}{E_{ab}^{2}}\left[\Im\left(J^{\alpha\beta}_{ab}J^{\gamma}_{ba}\right)+\Im\left(J^{\alpha\gamma}_{ab}J^{\beta}_{ba}\right)\right].
\end{align}
The first term is calculated as
\begin{equation}
\Im\left(J^{\beta}_{ab}J^{\gamma}_{ba}+J_{ba}^{\beta}J^{\gamma}_{ab}\right)=\Im\left(J^{\beta}_{ab}J^{\gamma}_{ba}+\left(J^{\beta}_{ab}J^{\gamma}_{ba}\right)^{*}\right)=0.
\end{equation}
From Eqs.~\eqref{eq:Im_J_J} and \eqref{eq:Delta}, the second term is evaluated as
\begin{align}
\left[\Delta^{\beta}_{ab}\Im\left(J^{\gamma}_{ba}J^{\alpha}_{ab}\right)\right]_{ab}  &=  -2(\partial_{\beta}\bm{g}\cdot\hat{\bm{g}})\left[(\partial_{\gamma}\bm{g}\times\partial_{\alpha}\bm{g})\cdot\hat{g}\right]
\begin{pmatrix}
0  &  1  \\
-1  &  0  
\end{pmatrix}
=0,
\end{align}
where we used $\bm{e}_{z}\parallel \hat{\bm{g}}$ and $(\partial_{x}\bm{g}\times\partial_{y}\bm{g})\cdot\hat{\bm{g}}=0$. The third term $\Delta^{\gamma}_{ab}\Im\left(J^{\beta}_{ba}J^{\alpha}_{ab}\right)$ also vanishes. From Eqs.~\eqref{eq:para} and \eqref{eq:diam}, the fourth term is calculated by
\begin{align}
\Im\left[J^{\alpha\beta}_{ab}J^{\gamma}_{ba}\right]_{ab}  &=  \left[(\partial_{\alpha\beta}\bm{g}\times\partial_{\gamma}\bm{g})\cdot\bm{e}_{z}\right]\begin{pmatrix}
0  &  1  \\
-1  &  0  \\
\end{pmatrix}_{ab}
=\left[(\partial_{\alpha\beta}\bm{g}\times\partial_{\gamma}\bm{g})\cdot\hat{\bm{g}}\right]
\begin{pmatrix}
0  &  1  \\
-1  &  0  \\
\end{pmatrix}_{ab},  \\
\Im\left[J^{\alpha\gamma}_{ab}J^{\beta}_{ba}\right]_{ab}  &=  \left[(\partial_{\alpha\gamma}\bm{g}\times\partial_{\beta}\bm{g})\cdot\bm{e}_{z}\right]
\begin{pmatrix}
0  &  1  \\
-1  &  0  \\
\end{pmatrix}_{ab}
=\left[(\partial_{\alpha\gamma}\bm{g}\times\partial_{\beta}\bm{g})\cdot\hat{\bm{g}}\right]
\begin{pmatrix}
0  &  1  \\
-1  &  0  \\
\end{pmatrix}_{ab}.
\end{align}
Thus, the shift current is obtained as
\begin{equation}
\sigma^{\alpha;\beta\gamma}_{\mathrm{shift}}=-\frac{\pi}{8}\sum_{\bm{k}}\frac{1}{g^{2}}\left[(\partial_{\alpha\beta}\bm{g}\times\partial_{\gamma}\bm{g})\cdot\hat{\bm{g}}+(\partial_{\alpha\gamma}\bm{g}\times\partial_{\beta}\bm{g})\cdot\hat{\bm{g}}\right]\Theta(g-\xi)\delta(\Omega-2g).
\end{equation}
Here, we differentiate the condition $(\partial_{\beta}\bm{g}\times\partial_{\gamma}\bm{g})\cdot\bm{g}=0$ by $\partial_{\alpha}$ and obtain
\begin{equation}
(\partial_{\alpha\beta}\bm{g}\times\partial_{\gamma}\bm{g})\cdot\bm{g}+(\partial_{\beta}\bm{g}\times\partial_{\alpha\gamma}\bm{g})\cdot\bm{g}+(\partial_{\beta}\bm{g}\times\partial_{\gamma}\bm{g})\cdot\partial_{\alpha}\bm{g}=0.
\end{equation}
In the two-dimensional system, the third term must vanish, $(\partial_{\beta}\bm{g}\times\partial_{\gamma}\bm{g})\cdot\partial_{\alpha}\bm{g}=0$. Using these results, we can rewrite the normal photocurrent conductivity as
\begin{equation}
\sigma^{\alpha;\beta\gamma}_{\mathrm{N}}=\sigma^{\alpha;\beta\gamma}_{\mathrm{shift}}=-\frac{\pi}{4}\sum_{\bm{k}}\frac{1}{g^{2}}[(\partial_{\alpha\beta}\bm{g}\times\partial_{\gamma}\bm{g})\cdot\hat{\bm{g}}]\Theta(g-\xi)\delta(\Omega-2g).
\end{equation}
This formula is shown in the main text.

Finally, we derive the electric injection current in the normal state when the normal Berry curvature factor is finite. From Eqs.~\eqref{eq:normal_Berry_curvature} and \eqref{eq:Delta}, the electric injection current is given by
\begin{align}
\sigma^{\alpha;\beta\gamma}_{\mathrm{Einj}} &= -\frac{i\pi}{4\eta}\sum_{a\neq b}(J^{\alpha}_{aa}-J^{\alpha}_{bb})\Omega^{\beta\gamma}_{ba}f_{ab}\delta(\Omega-E_{ba}),  \\
&=-\frac{i\pi}{4\eta}\sum_{\bm{k}}(\partial_{\alpha}\bm{g} \cdot\hat{\bm{g}})\frac{[(\partial_{\beta}\bm{g}\times\partial_{\gamma}\bm{g})\cdot\hat{\bm{g}}]}{g^{2}}\Theta(g-\xi)\delta(\Omega-2g),
\end{align}
where we use $E_{ba}=\Omega\geq 0$ due to $\delta(\Omega-E_{ba})$.

\section{Analytical solutions of electric injection current in the superconducting state}

In this section, we derive the electric injection current of the $\mathcal{T}$ symmetric superconductor under various conditions on the pair potential. The BdG Hamiltonian is given by
\begin{equation}
H_{\mathrm{BdG}}(\bm{k}) =
\begin{pmatrix}
H_{\rm N}(\bm{k}) & \Delta(\bm{k}) \\
\Delta^{\dagger}(\bm{k}) & -\left[H_{\rm N}(-\bm{k})\right]^{T} \\
\end{pmatrix},
\end{equation}
where $H_{\mathrm{N}}(\bm{k})= \xi_{\bm{k}}+\bm{g_{k}}\cdot\bm{\sigma}$ is the Hamiltonian in the normal state, and $\Delta(\bm{k})=[\psi_{\bm{k}}+\bm{d_{k}}\cdot\bm{\sigma}]i\sigma_{y}$ is the pair potential of superconductivity. Under appropriate gauge, $\psi_{\bm k}$ and $\bm{d_{k}}$ are real valued due to the $\mathcal{T}$ symmetry.
The BdG Hamiltonian in the coordinate $\hat{z}\parallel\hat{g}$ has the form
\begin{equation}
H_{\mathrm{BdG}}(\bm{k})= \begin{pmatrix}
\xi+g          & 0            & -d_{x}+id_{y}  & \psi+d_{z} \\
0              & \xi-g        & -\psi+d_{z}    & d_{x}+id_{y} \\
-d_{x}-id_{y}  & -\psi+d_{z}  & -\xi+g         & 0\\
\psi+d_{z}     & d_{x}-id_{y} & 0              &-\xi-g
\end{pmatrix},
\end{equation}
where indices of $\bm{d}$ are based on the new coordinates. We see that $\psi$ and $d_{z}$ represent intraband pairing, while $d_{x}$ and $d_{y}$ are interband pairing.
Electric injection current is given by 
\begin{equation}
\label{eq:Einj_formula}
\sigma^{\alpha;\beta\gamma}_{\mathrm{Einj}} = -\frac{i\pi}{8\eta}\sum_{a\neq b}(J^{\alpha}_{aa}-J^{\alpha}_{bb})\Omega^{\lambda_{\beta}\lambda_{\gamma}}_{ba}f_{ab}\delta(\Omega-E_{ba}), 
\end{equation}
where the Berry curvature is defined as 
\begin{equation}
\Omega^{\lambda_{\alpha}\lambda_{\beta}}_{ab} = -\frac{2}{E^{2}_{ab}}\Im[J^{\alpha}_{ab}J^{\beta}_{ba}].
\end{equation}
The paramagnetic current density operator is obtained as
\begin{equation}
J^{\alpha}=
\begin{pmatrix}
\partial_{\alpha}\xi+\partial_{\alpha}g_{z}     & \partial_{\alpha}g_{x}-i\partial_{\alpha}g_{y}  & 0                                                & 0 \\
\partial_{\alpha}g_{x}+i\partial_{\alpha}g_{y}  & \partial_{\alpha}\xi-\partial_{\alpha}g_{z}     & 0                                                & 0 \\
0                                               & 0                                               & \partial_{\alpha}\xi-\partial_{\alpha}g_{z}      & -\partial_{\alpha}g_{x}-i\partial_{\alpha}g_{y} \\
0                                               & 0                                               & -\partial_{\alpha}g_{x}+i\partial_{\alpha}g_{y}  & \partial_{\alpha}\xi+\partial_{\alpha}g_{z}
\end{pmatrix}.
\end{equation}
We derive the electric injection current under the condition of vanishing interband pairing ($d_{x}=d_{y}=0$) and under the condition of vanishing intraband pairing ($\psi=d_{z}=0$) in the following subsections.

\subsection{Zero interband pairing condition}
First, we calculate the electric injection current of superconductors in the absence of the interband pairing. By an appropriate permutation of the bases, the BdG Hamiltonian and paramagnetic current density operator are rewritten as
\begin{equation}
H_{\mathrm{BdG}}(\bm{k})=
\begin{pmatrix}
\xi+g       & \psi+d_{z}  & 0            & 0  \\
\psi+d_{z}  & -\xi-g      & 0            & 0  \\
0           & 0           & \xi-g        & -\psi+d_{z}  \\
0           & 0           & -\psi+d_{z}  & -\xi+g  \\
\end{pmatrix},
\end{equation}
\begin{equation}
J^{\alpha}=
\begin{pmatrix}
\partial_{\alpha}\xi+\partial_{\alpha}g_{z}  & 0  & \partial_{\alpha}g_{x}-i\partial_{\alpha}g_{y} & 0 \\
0  & \partial_{\alpha}\xi+\partial_{\alpha}g_{z}  & 0  & -\partial_{\alpha}g_{x}+i\partial_{\alpha}g_{y} \\
\partial_{\alpha}g_{x}+i\partial_{\alpha}g_{y}  & 0  & \partial_{\alpha}\xi-\partial_{\alpha}g_{z}  & 0  \\
0  & -\partial_{\alpha}g_{x}-i\partial_{\alpha}g_{y}  & 0  & \partial_{\alpha}\xi-\partial_{\alpha}g_{z}  \\
\end{pmatrix}.
\end{equation}
We can diagonalize the diagonal block of the BdG Hamiltonian
\begin{align}
U^{\dagger}
\begin{pmatrix}
\xi+g       & \psi+d_{z}  \\
\psi+d_{z}  & -\xi-g
\end{pmatrix}
U = \sqrt{(\xi+g)^{2}+(\psi+d_{z})^{2}}\sigma_{z}=u\sigma_{z}, \\
V^{\dagger}
\begin{pmatrix}
\xi-g        & -\psi+d_{z}  \\
-\psi+d_{z}  & -\xi+g       \\
\end{pmatrix}
V = \sqrt{(\xi-g)^{2}+(\psi-d_{z})^{2}}\sigma_{z}=v\sigma_{z},
\end{align}
by using unitary operators
\begin{align}
U=\frac{1}{\sqrt{2u(u+u_{z})}}
\begin{pmatrix}
u_{x} \\
u_{y} \\
u_{z}+u 
\end{pmatrix}
\cdot\bm{\sigma} ,\quad \bm{u}=
\begin{pmatrix}
\psi+d_{z} \\
0 \\
\xi+g 
\end{pmatrix},\\
V = \frac{1}{\sqrt{2v(v+v_{z})}}
\begin{pmatrix}
v_{x} \\
v_{y} \\
v+v_{z} \\
\end{pmatrix}
\cdot\bm{\sigma}, \quad \bm{v}=
\begin{pmatrix}
-\psi+d_{z} \\
0 \\
\xi-g \\
\end{pmatrix}.
\end{align}
We carry out the unitary transformation of the paramagnetic current density operator
\begin{equation}
U_{\mathrm{tot}}^{\dagger}J^{\alpha}U_{\mathrm{tot}}=
\begin{pmatrix}
U^{\dagger} & 0  \\
0           & V^{\dagger} \\
\end{pmatrix}
\begin{pmatrix}
A^{\alpha}  & B^{\alpha}  \\
(B^{\alpha})^{\dagger} & C^{\alpha}
\end{pmatrix}
\begin{pmatrix}
U  & 0  \\
0  & V  
\end{pmatrix}=
\begin{pmatrix}
U^{\dagger}A^{\alpha}U  & U^{\dagger}B^{\alpha}V  \\
V^{\dagger}(B^{\alpha})^{\dagger}U  & V^{\dagger}C^{\alpha}V
\end{pmatrix},
\end{equation}
where $A$, $B$, and $C$ are defined as
\begin{equation}
A^{\alpha}=(\partial_{\alpha}\xi+\partial_{\alpha}g_{z})\sigma_{0}, \quad B^{\alpha}=(\partial_{\alpha}g_{x}-i\partial_{\alpha}g_{y})\sigma_{z}, \quad C^{\alpha}=(\partial_{\alpha}\xi-\partial_{\alpha}g_{z})\sigma_{0}.
\end{equation}
The off-diagonal block of the transformed current density operator is obtained as
\begin{equation}
\label{eq:b_tilde_comp}
U^{\dagger} B^{\alpha} V=\frac{\partial_{\alpha}g_{x}-i\partial_{\alpha}g_{y}}{2\sqrt{uv(u+u_{z})(v+v_{z})}}
\begin{pmatrix}
(v+v_{z})u_{x}+(u+u_{z})v_{x}  \\
0  \\
uv+u_{z}v+v_{z}u+u_{z}v_{z}-u_{x}v_{x}
\end{pmatrix}
\cdot\bm{\sigma} \equiv \tilde{B}^{\alpha}.
\end{equation}
Since $A$ and $C$ are invariant under unitary operation,  we obtain $\tilde{A} \equiv U^{\dagger}AU=A$ and $\tilde{C} \equiv V^{\dagger}CV=C$.
The factors $\Im[J^{\alpha}_{ab}J^{\beta}_{ba}]$ of the Berry curvature are given in the matrix representation by
\begin{align}
\label{eq:gene_Berry_curv}
&\left(\Im[J^{\alpha}_{ab}J^{\beta}_{ba}]\right)_{ab} \notag \\
&=\begin{pmatrix}
0  &  \tilde{\bm{a}}^{\alpha}\times\tilde{\bm{a}}^{\beta}\cdot\bm{e}_{z}  &  \Im\left[(\tilde{b}^{\alpha}_{0}+\tilde{b}^{\alpha}_{z})(\tilde{b}^{\beta*}_{0}+\tilde{b}^{\beta*}_{z})\right]  &  \Im\left[(\tilde{b}^{\alpha}_{x}-i\tilde{b}^{\alpha}_{y})(\tilde{b}^{\beta*}_{x}+i\tilde{b}^{\beta*}_{y})\right] \\
-\tilde{\bm{a}}^{\alpha}\times\tilde{\bm{a}}^{\beta}\cdot\bm{e}_{z}  &  0  &  \Im\left[(\tilde{b}^{\alpha}_{x}+i\tilde{b}^{\alpha}_{y})(\tilde{b}^{\beta*}_{x}-i\tilde{b}^{\beta*}_{y})\right]  &  \Im\left[(\tilde{b}^{\alpha}_{0}-\tilde{b}^{\alpha}_{z})(\tilde{b}^{\beta*}_{0}-\tilde{b}^{\beta*}_{z})\right]  \\
-\Im\left[(\tilde{b}^{\alpha}_{0}+\tilde{b}^{\alpha}_{z})(\tilde{b}^{\beta*}_{0}+\tilde{b}^{\beta*}_{z})\right]  &  -\Im\left[(\tilde{b}^{\alpha}_{x}+i\tilde{b}^{\alpha}_{y})(\tilde{b}^{\beta*}_{x}-i\tilde{b}^{\beta*}_{y})\right]  & 0  &  \tilde{\bm{c}}^{\alpha}\times\tilde{\bm{c}}^{\beta}\cdot\bm{e}_{z}  \\
-\Im\left[(\tilde{b}^{\alpha}_{x}-i\tilde{b}^{\alpha}_{y})(\tilde{b}^{\beta*}_{x}+i\tilde{b}^{\beta*}_{y})\right]  &  -\Im\left[(\tilde{b}^{\alpha}_{0}-\tilde{b}^{\alpha}_{z})(\tilde{b}^{\beta*}_{0}-\tilde{b}^{\beta*}_{z})\right]  &  -\tilde{\bm{c}}^{\alpha}\times\tilde{\bm{c}}^{\beta}\cdot\bm{e}_{z}  &  0  
\end{pmatrix},
\end{align}
where we use the representation $\tilde{A}=\tilde{a}_{0}+\tilde{\bm{a}}\cdot\bm{\sigma}$, $\tilde{B}=\tilde{b}_{0}+\tilde{\bm{b}}\cdot\bm{\sigma}$, and $\tilde{C}=\tilde{c}_{0}+\tilde{\bm{c}}\cdot\bm{\sigma}$. The matrix element $\tilde{b}^{\alpha}_{x}-i\tilde{b}^{\alpha}_{y}$ of $J^{\alpha}$ is obtained from Eq.~\eqref{eq:b_tilde_comp},
\begin{equation}
\tilde{b}^{\alpha}_{x}-i\tilde{b}^{\alpha}_{y}=\tilde{b}^{\alpha}_{x}=\frac{(\partial_{\alpha}g_{x}-i\partial_{\alpha}g_{y})[(v+v_{z})u_{x}+(u+u_{z})v_{x}]}{2\sqrt{uv(u+u_{z})(v+v_{z})}},
\end{equation}
and we obtain
\begin{equation}
\label{eq:berry_b_x_intra}
\Im\left[(\tilde{b}^{\alpha}_{x}-i\tilde{b}^{\alpha}_{y})(\tilde{b}^{\beta*}_{x}+i\tilde{b}^{\beta*}_{y})\right]=\frac{[(v+v_{z})u_{x}+(u+u_{z})v_{x}]^{2}}{4uv(u+u_{z})(v+v_{z})}[(\partial_{\alpha}\bm{g}\times\partial_{\beta}\bm{g})\cdot\bm{e}_{z}].
\end{equation}
The diagonal block of Eq.~\eqref{eq:gene_Berry_curv} vanishes because $\tilde{\bm{a}}=\tilde{\bm{c}}=0$. From $\tilde{A}=(\partial_{\alpha}\xi+\partial_{\alpha}g_{z})\sigma_{0}$ and $\tilde{C}=(\partial_{\alpha}\xi-\partial_{\alpha}g_{z})\sigma_{0}$, the velocity difference ($J^{\alpha}_{aa}-J^{\alpha}_{bb})$ is given by
\begin{equation}
\label{eq:diff_velocity}
\left(J^{\alpha}_{aa}-J^{\alpha}_{bb}\right)_{ab}=
\begin{pmatrix}
0  &  0  &  2\partial_{\alpha}g_{z}  &  2\partial_{\alpha}g_{z}  \\
0  &  0  &  2\partial_{\alpha}g_{z}  &  2\partial_{\alpha}g_{z}  \\
-2\partial_{\alpha}g_{z}  &  -2\partial_{\alpha}g_{z}  &  0  &  0  \\
-2\partial_{\alpha}g_{z}  &  -2\partial_{\alpha}g_{z}  &  0  &  0  \\
\end{pmatrix}.
\end{equation}
Operating the unitary transformation, we obtain the diagonalized BdG Hamiltonian $\tilde{H}_{\mathrm{BdG}}(\bm{k})=\mathrm{diag}(u, -u, v, -v)$. Taking the zero temperature limit, we obtain the formula of the injection current,
\begin{align}
\sigma^{\alpha;\beta\gamma}_{\mathrm{Einj}} &= \frac{i\pi}{4\eta}\sum_{\bm{k}}\sum_{a\neq b}(J^{\alpha}_{aa}-J^{\alpha}_{bb})\frac{\Im[J^{\beta}_{ba}J^{\gamma}_{ab}]}{E_{ab}^{2}}f_{ab}\delta(\Omega-E_{ba}),  \\
&=-\frac{i\pi}{4\eta}\sum_{\bm{k}}\frac{[(u+u_{z})v_{x}+(v+v_{z})u_{x}]^{2}}{uv(u+u_{z})(v+v_{z})(u+v)^{2}}(\partial_{\alpha}g_{z})\left[(\partial_{\beta}\bm{g}\times\partial_{\gamma}\bm{g})\cdot\bm{e}_{z}\right]\delta(\Omega-(u+v)),
\end{align}
where we used Eqs.~\eqref{eq:Einj_formula}, \eqref{eq:gene_Berry_curv}, \eqref{eq:berry_b_x_intra}, and \eqref{eq:diff_velocity}. We fixed the $\hat{z}\parallel\hat{g}$ coordinate, and then we can rewrite 
\begin{equation}
\sigma^{\alpha;\beta\gamma}_{\mathrm{Einj}} = -\frac{i\pi}{4\eta}\sum_{\bm{k}}\frac{\left[F_{+}(-\psi_{\bm{k}}+d_{\bm{k}})+F_{-}(\psi_{\bm{k}}+d_{\bm{k}})\right]^{2}}{E_{+}E_{-}F_{+}F_{-}(E_{+}+E_{-})^{2}}(\partial_{\alpha}\bm{g_{k}}\cdot\hat{\bm{g_{k}}})\left[(\partial_{\beta}\bm{g_{k}}\times\partial_{\gamma}\bm{g_{k}})\cdot\hat{\bm{g}}_{\bm k}\right]\delta\left(\Omega-(E_{+}+E_{-})\right),
\end{equation}
where we defined 
\begin{align}
E_{+} &=u=\sqrt{(\xi_{\bm{k}} + g_{\bm{k}})^{2}+(\psi_{\bm{k}}+\bm{d_{k}}\cdot\hat{\bm{g}}_{\bm{k}})},  \\
E_{-} &=u=\sqrt{(\xi_{\bm{k}} - g_{\bm{k}})^{2}+(\psi_{\bm{k}}-\bm{d_{k}}\cdot\hat{\bm{g}}_{\bm{k}})},  \\
F_{\pm} &= E_{\pm} + \xi_{\bm{k}} \pm g_{\bm{k}}.
\end{align}

\subsection{Zero intraband pairing condition}
Next, we derive the electric injection current of superconductors in the absence of intraband pairing. By an appropriate permutation of the bases, the BdG Hamiltonian and paramagnetic current density operator are given by
\begin{equation}
H_{\mathrm{BdG}}(\bm{k})=
\begin{pmatrix}
\xi+g          & -d_{x}+id_{y}  & 0             & 0  \\
-d_{x}-id_{y}  & -\xi+g         & 0             & 0  \\
0              & 0              & -\xi-g        & d_{x}-id_{y}  \\
0              & 0              & d_{x}+id_{y}  & \xi-g  \\
\end{pmatrix}=
\begin{pmatrix}
h  &  0  \\
0  &  -h
\end{pmatrix},
\end{equation}
\begin{equation}
J^{\alpha}=
\begin{pmatrix}
\partial_{\alpha}\xi+\partial_{\alpha}g_{z}  & 0  & 0 & \partial_{\alpha}g_{x}-i\partial_{\alpha}g_{y} \\
0  & \partial_{\alpha}\xi-\partial_{\alpha}g_{z}  & -\partial_{\alpha}g_{x}-i\partial_{\alpha}g_{y}  & 0 \\
0  & -\partial_{\alpha}g_{x}+i\partial_{\alpha}g_{y}  & \partial_{\alpha}\xi+\partial_{\alpha}g_{z}  & 0  \\
\partial_{\alpha}g_{x}+i\partial_{\alpha}g_{y}  & 0 & 0  & \partial_{\alpha}\xi-\partial_{\alpha}g_{z}  \\
\end{pmatrix}.
\end{equation}
We can diagonalize the BdG Hamiltonian
\begin{equation}
\begin{pmatrix}
U^{\dagger}  &  0  \\
0            &  U^{\dagger}  \\
\end{pmatrix}
\begin{pmatrix}
h  &  0  \\
0  &  -h \\
\end{pmatrix}
\begin{pmatrix}
U  &  0  \\
0  &  U  \\
\end{pmatrix}
=\mathrm{diag}\left(g+\sqrt{\xi^{2}+d^{2}}, g-\sqrt{\xi^{2}+d^{2}}, -g-\sqrt{\xi^{2}+d^{2}}, -g+\sqrt{\xi^{2}+d^{2}}\right),
\end{equation}
where we define the unitary operator $U$ as
\begin{equation}
U = \frac{1}{\sqrt{2u(u+u_{z})}}
\begin{pmatrix}
u_{x}  \\
u_{y}  \\
u+u_{z} 
\end{pmatrix}
\cdot\bm{\sigma}, \quad {\bm u}=
\begin{pmatrix}
-d_{x}  \\
-d_{y}  \\
\xi
\end{pmatrix}.
\end{equation}
We carry out the unitary transformation of the paramagnetic current density operator
\begin{equation}
U_{\mathrm{tot}}^{\dagger}J^{\alpha}U_{\mathrm{tot}}=
\begin{pmatrix}
U^{\dagger} & 0  \\
0           & U^{\dagger} \\
\end{pmatrix}
\begin{pmatrix}
A^{\alpha}  & B^{\alpha}  \\
(B^{\alpha})^{\dagger} & A^{\alpha}
\end{pmatrix}
\begin{pmatrix}
U  & 0  \\
0  & U  
\end{pmatrix}=
\begin{pmatrix}
U^{\dagger}A^{\alpha}U  & U^{\dagger}B^{\alpha}U  \\
U^{\dagger}(B^{\alpha})^{\dagger}U  & U^{\dagger}A^{\alpha}U
\end{pmatrix},
\end{equation}
where $A$ and $B$ are defined as
\begin{equation}
A=
\begin{pmatrix}
\partial_{\alpha}\xi+\partial_{\alpha}g_{z}  &  0  \\
0  &  \partial_{\alpha}\xi-\partial_{\alpha}g_{z}
\end{pmatrix}
, \quad B=
\begin{pmatrix}
0  &  \partial_{\alpha}g_{x}-i\partial_{\alpha}g_{y}  \\
-\partial_{\alpha}g_{x}-i\partial_{\alpha}g_{y}  &  0 
\end{pmatrix}.
\end{equation}
We obtain the off-diagonal block of the transformed current density operator as 
\begin{equation}
\label{eq:b_comp_inter}
\tilde{B}^{\alpha}=U^{\dagger}B^{\alpha}U=\tilde{\bm{b}}\cdot\bm{\sigma}, \quad \tilde{\bm{b}}^{\alpha}=
\begin{pmatrix}
i\partial_{\alpha}g_{y}  \\
-i\partial_{\alpha}g_{x}  \\
0
\end{pmatrix}  +  \frac{i(\bm{d}\times\partial_{\alpha}\bm{g})_{z}}{u(u+\xi)}
\begin{pmatrix}
-d_{x}  \\
-d_{y}  \\
u+\xi  
\end{pmatrix}.
\end{equation}
The matrix element $\tilde{b}^{\alpha}_{x}-i\tilde{b}^{\alpha}_{y}$ is given by Eq.~\eqref{eq:b_comp_inter},
\begin{equation}
\tilde{b}^{\alpha}_{x}-i\tilde{b}^{\alpha}_{y}=-(\partial_{\alpha}g_{x}-i\partial_{\alpha}g_{y})-\frac{i(\bm{d}\times\partial_{\alpha}\bm{g})_{z}}{u(u+\xi)}(d_{x}-id_{y}),
\end{equation}
and we obtain
\begin{equation}
\Im\left[(\tilde{b}^{\alpha}_{x}-i\tilde{b}^{\alpha}_{y})(\tilde{b}^{\beta*}_{x}+i\tilde{b}^{\beta*}_{y})\right] = \left(1-\frac{d_{x}^{2}+d_{y}^{2}}{u(u+\xi)}\right)\left[(\partial_{\alpha}\bm{g}\times\partial_{\beta}\bm{g})\cdot\bm{e}_{z}\right].
\end{equation}
Since $\tilde{b}^{\alpha}_{x}$ and $\tilde{b}^{\alpha}_{y}$ are pure imaginary, the relation between components of the Berry curvature is obtained as
\begin{align}
\label{eq:inter_berry}
\Im\left[(\tilde{b}^{\alpha}_{x}+i\tilde{b}^{\alpha}_{y})(\tilde{b}^{\beta*}_{x}-i\tilde{b}^{\beta*}_{y})\right] &= -\Im\left[(\tilde{b}^{\alpha*}_{x}-i\tilde{b}^{\alpha*}_{y})(\tilde{b}^{\beta}_{x}+i\tilde{b}^{\beta}_{y})\right], \notag \\
& =-\Im\left[(\tilde{b}^{\alpha}_{x}-i\tilde{b}^{\alpha}_{y})(\tilde{b}^{\beta*}_{x}+i\tilde{b}^{\beta*}_{y})\right], \notag \\
& =-\left(1-\frac{d_{x}^{2}+d_{y}^{2}}{u(u+\xi)}\right)\left[(\partial_{\alpha}\bm{g}\times\partial_{\beta}\bm{g})\cdot\bm{e}_{z}\right].
\end{align}
The diagonal block of the transformed current density operator is
\begin{equation}
\label{eq:tilde_A}
\tilde{A}^{\alpha}=U^{\dagger}A^{\alpha}U=a_{0}^{\alpha}+\tilde{\bm{a}}^{\alpha}\cdot\bm{\sigma}, \quad a_{0}^{\alpha}=\partial_{\alpha}\xi, \quad \tilde{\bm{a}}^{\alpha}=
\begin{pmatrix}
0   \\
0   \\
-\partial_{\alpha}g_{z} \\
\end{pmatrix} + \frac{\partial_{\alpha}g_{z}}{u}
\begin{pmatrix}
-d_{x}  \\
-d_{y}  \\
u+\xi  \\
\end{pmatrix}.
\end{equation}
A component $\tilde{\bm{a}}^{\alpha}\times\tilde{\bm{a}}^{\beta}\cdot\bm{e}_{z}$ in Eq.~\eqref{eq:gene_Berry_curv} is zero. 
From Eq.~\eqref{eq:tilde_A}, the group velocity difference $(J_{aa}^{\alpha}-J_{bb}^{\beta})$ is given by
\begin{equation}
\label{eq:inter_diff_velo}
\left(J^{\alpha}_{aa}-J^{\alpha}_{bb}\right)_{ab}=\frac{\xi\partial_{\alpha}g_{z}}{u}\times
\begin{pmatrix}
0  &  1  &  0  &  1  \\
-1  &  0  &  -1  &  0  \\
0  &  1  &  0  &  1  \\
-1  &  0  &  -1  &  0  \\
\end{pmatrix}.
\end{equation}
Taking the zero temperature limit, we obtain
\begin{align}
\sigma^{\alpha;\beta\gamma}_{\mathrm{Einj}} &= \frac{i\pi}{4\eta}\sum_{\bm{k}}\sum_{a\neq b}(J^{\alpha}_{aa}-J^{\alpha}_{bb})\frac{\Im[J^{\beta}_{ba}J^{\gamma}_{ab}]}{E_{ab}^{2}}f_{ab}\delta(\Omega-E_{ba}),  \\
&=-\frac{i\pi}{4\eta}\sum_{\bm{k}} \frac{\xi\partial_{\alpha}g_{z}}{ug^{2}}\left(1-\frac{d_{x}^{2}+d_{y}^{2}}{u(u+\xi)}\right) \left[(\partial_{\beta}\bm{g}\times\partial_{\gamma}\bm{g})\cdot\bm{e}_{z}\right]\Theta\left(g-\sqrt{\xi^{2}+d^{2}}\right)\delta\left(\Omega-2g\right),
\end{align}
where we used Eqs.~\eqref{eq:Einj_formula}, \eqref{eq:gene_Berry_curv}, \eqref{eq:inter_berry}, and \eqref{eq:inter_diff_velo}, and $\Theta$ is a step function. We fixed the $\hat{z}\parallel\hat{g}$ coordinate, and we can rewrite
\begin{align}
\sigma^{\alpha;\beta\gamma}_{\mathrm{Einj}}=\frac{-i\pi}{4\eta}&\sum_{\bm{k}}\frac{\xi_{\bm{k}}(\partial_{\alpha}\bm{g_{k}}\cdot\hat{\bm{g}}_{\bm{k}})}{u_{\bm{k}}g_{\bm{k}}^{2}}\left(1-\frac{d_{\bm{k}}^{2}}{u_{\bm{k}}(u_{\bm{k}}+\xi_{\bm{k}})}\right)\left[\left(\partial_{\beta}\bm{g_{k}}\times\partial_{\gamma}\bm{g_{k}}\right)\cdot\hat{\bm{g}}_{\bm{k}}\right]\Theta\left(g_{\bm{k}}-\sqrt{\xi_{\bm{k}}^{2} + d_{\bm{k}}^{2}}\right)\delta\left(\Omega-2 g_{\bm{k}}\right),
\end{align}
where we defined $u_{\bm{k}}$ as
\begin{equation}
u_{\bm{k}}=\sqrt{\xi_{\bm{k}}^{2}+d_{\bm{k}}^{2}}.
\end{equation}

\subsection{Approximate formula for electric injection current}
Finally, we derive an approximate formula for the electric injection current when both intraband and interband pairing are present. First, we approximate the eigenstates and eigenvalues of the Hamiltonian. We decompose the normal part of BdG Hamiltonian
\begin{align}
H_{\mathrm{N}}(\bm{k})=\xi_{\bm{k}}+\bm{g}_{\parallel\bm{k}}+\bm{g}_{\perp\bm{k}}=\tilde{H}_{\mathrm{N}}+\bm{g}_{\perp\bm{k}},
\end{align}
where the decomposition of $\bm{g_{k}}$ is performed as
\begin{align}
\bm{g_{k}}=\bm{g}_{\parallel\bm{k}}+\bm{g}_{\perp\bm{k}}, \quad \bm{g}_{\parallel\bm{k}}\times\bm{d_{k}}=0,  \quad   \bm{g}_{\perp\bm{k}}\cdot\bm{d_{k}}=0.
\end{align}
The approximate BdG Hamiltonian $\tilde{H}_{\mathrm{BdG}}$ is defined by 
\begin{align}
\tilde{H}_{\mathrm{BdG}}(\bm{k})=
\begin{pmatrix}
\tilde{H}_{\mathrm{N}}(\bm{k})  &  \Delta(\bm{k})  \\
\Delta^{\dagger}(\bm{k})  &  -\left[\tilde{H}_{\mathrm{N}}(-\bm{k})\right]^{\mathrm{T}}
\end{pmatrix}.
\end{align}
We obtain the eigenvalues $\tilde{E}_{a}$ and eigenvectors $\ket{\tilde{a}_{\lambda}}$ of $\tilde{H}_{\mathrm{BdG}}$ and approximate the velocity operator $J^{\alpha}_{ab}(\bm{k})$
\begin{align}
\tilde{J}^{\alpha}_{ab}(\bm{k})=-\lim_{\bm{\lambda}\rightarrow\bm{0}}\Braket{\tilde{a}_{\lambda}|\frac{\partial H_{\lambda}}{\partial \lambda_{a}}|\tilde{b}_{\lambda}}.
\end{align}
Here, we approximately evaluate the injection current $\sigma^{\alpha;\beta\gamma}_{\mathrm{Einj}}$ by using the eigenvalues $\tilde{E}_{a}$ and eigenvectors $\ket{\tilde{a}_{\lambda}}$.
\begin{align}
\label{eq:appro_Einj}
\sigma^{\alpha;\beta\gamma}_{\mathrm{Einj}}\simeq\frac{-\pi}{4\eta}\sum_{\bm{k}}&\sum_{a,b}\left[\tilde{J}^{\alpha}_{aa}(\bm{k})-\tilde{J}^{\alpha}_{bb}(\bm{k})\right]\frac{\tilde{J}^{\beta}_{ba}(\bm{k})\tilde{J}^{\gamma}_{ab}(\bm{k})}{{\tilde{E}_{ba}}^{2}}\tilde{f}_{ab}\delta\left(\Omega-\tilde{E}_{ab}\right),
\end{align}
where $\tilde{f}_{ab}$ is the difference of Fermi-Dirac distributions between the eigenvalues $\tilde{E}_{a}$ and $\tilde{E}_{b}$. Equation~\eqref{eq:appro_Einj} is almost the same as the electric injection current in the case of vanishing interband pairing. Indeed, we obtain  
\begin{align}
\sigma^{\alpha;\beta\gamma}_{\mathrm{Einj}} = -\frac{i\pi}{4\eta}\sum_{\bm{k}}\frac{\left[F_{+}(-\psi_{\bm{k}}+d_{\bm{k}})+F_{-}(\psi_{\bm{k}}+d_{\bm{k}})\right]^{2}}{E_{+}E_{-}F_{+}F_{-}(E_{+}+E_{-})^{2}}(\partial_{\alpha}\bm{g_{k}}\cdot\hat{\bm{g}}_{\parallel\bm{k}})\left[(\partial_{\beta}\bm{g_{k}}\times\partial_{\gamma}\bm{g_{k}})\cdot\hat{\bm{g}}_{\parallel\bm{k}}\right]\delta\left(\Omega-(E_{+}+E_{-})\right),
\end{align}
where $E_{\pm}$ and $F_{\pm}$ are defined as
\begin{align}
E_{+} &=u=\sqrt{(\xi_{\bm{k}} + g_{\parallel\bm{k}})^{2}+(\psi_{\bm{k}}+\bm{d_{k}}\cdot\hat{\bm{g}}_{\parallel\bm{k}})},  \\
E_{-} &=v=\sqrt{(\xi_{\bm{k}} - g_{\parallel\bm{k}})^{2}+(\psi_{\bm{k}}-\bm{d_{k}}\cdot\hat{\bm{g}}_{\parallel\bm{k}})},  \\
F_{\pm} &= E_{\pm} + \xi_{\bm{k}} \pm g_{\parallel\bm{k}}.
\end{align}
Using the relationship that $\bm{g}_{\parallel\bm{k}}$ is parallel to $\bm{d_{k}}$, we finally obtain 
\begin{align}
\sigma^{\alpha;\beta\gamma}_{\mathrm{Einj}} = -\frac{i\pi}{4\eta}\sum_{\bm{k}}\frac{\left[F_{+}(-\psi_{\bm{k}}+d_{\bm{k}})+F_{-}(\psi_{\bm{k}}+d_{\bm{k}})\right]^{2}}{E_{+}E_{-}F_{+}F_{-}(E_{+}+E_{-})^{2}}(\partial_{\alpha}\bm{g_{k}}\cdot\hat{\bm{d}}_{\bm{k}})\left[(\partial_{\beta}\bm{g_{k}}\times\partial_{\gamma}\bm{g_{k}})\cdot\hat{\bm{d}}_{\bm{k}}\right]\delta\left(\Omega-(E_{+}+E_{-})\right),
\end{align}
where $E_{\pm}$ and $F_{\pm}$ are obtained as
\begin{align}
E_{\pm} &=\sqrt{(\xi_{\bm{k}} \pm \bm{g_{k}}\cdot\hat{\bm{d}}_{\bm{k}})^{2}+(\psi_{\bm{k}} \pm d_{\bm{k}})},  \\
F_{\pm} &= E_{\pm} + \xi_{\bm{k}} \pm \bm{g_{k}}\cdot\hat{\bm{d}}_{\bm{k}}.
\end{align}

\bibliographystyle{apsrev4-1}
\bibliography{reference}

\end{document}